\documentclass[letterpaper, 10 pt, conference]{ieeeconf}

\IEEEoverridecommandlockouts
\usepackage[latin1]{inputenc}
\usepackage[english]{babel}
\usepackage{amsmath,amssymb}
\usepackage{graphicx,graphicx,verbatim}
\usepackage{verbatim}
\usepackage{color}

\newtheorem{thm}{Theorem}
\newtheorem{cor}{Corollary}
\newtheorem{lem}{Lemma} 

\newtheorem{defi}{Definition}
\newtheorem{prop}{Proposition}

\newtheorem{exx}{Example}
\newtheorem{remm}{Remark}

\newcommand\real{\ensuremath{{\mathbb R}}}
\newcommand\realn{\ensuremath{{\mathbb{R}^n}}}

\newcommand\mymatrix[2]{\left[\begin{array}{#1} #2 \end{array}\right]}

\newcommand{\calC}{\mathcal{C}}

\newcommand{\calH}{\mathcal{H}}

\newcommand{\calU}{\mathcal{U}}
\newcommand{\calV}{\mathcal{V}}

\newcommand{\calK}{\mathcal{K}}
\newcommand{\calW}{\mathcal{W}}

\newcommand{\calX}{\mathcal{X}}

\begin{document}
\title{\LARGE \bf Differential analysis of nonlinear systems: \\ revisiting the pendulum example}

\author{F. Forni, R. Sepulchre
\thanks{
F. Forni and R. Sepulchre are with the University of Cambridge, Department of Engineering, 
Trumpington Street, Cambridge CB2 1PZ, and with the Department of Electrical Engineering and Computer Science, 
University of Li{\`e}ge, 4000 Li{\`e}ge, Belgium, \texttt{ff286@cam.ac.uk|r.sepulchre@eng.cam.ac.uk}.
The research is supported by FNRS.
The paper presents research results of the Belgian Network DYSCO
(Dynamical Systems, Control, and Optimization), funded by the
Interuniversity Attraction Poles Programme, initiated by the Belgian
State, Science Policy Office. The scientific responsibility rests with
its authors.} 
}

\date{\today}

\maketitle

\begin{abstract}  
Differential analysis aims at inferring global properties of nonlinear behaviors
from the local analysis of the linearized dynamics. The paper motivates and
illustrates the use of differential analysis on the nonlinear pendulum model,
an archetype example of nonlinear behavior. Special emphasis is put on recent
work by the authors in this area, which includes a differential Lyapunov framework
for contraction analysis \cite{Forni2014},  and the concept of differential positivity \cite{Forni2014a_ver1}.
\end{abstract}

\section{Introduction}

The purpose of this tutorial paper is to revisit the role of linearization in nonlinear
systems analysis and to present recent developments of this {\it differential} approach
to systems and control theory. Linearization is often considered as a synonym of 
{\it local} analysis, whereas nonlinear
systems analysis aims at a global understanding of the system behavior. The focus of the paper is
therefore on system properties that allow to address non-local questions through
the local-in-nature analysis of a differential approach. Such properties have been sporadically 
studied in the control community, perhaps most importantly through the contraction property
advocated in the seminal paper of Lohmiller and Slotine \cite{Lohmiller1998}, but 
they play at best a secondary role in the main textbooks of nonlinear control. While it
is not the aim of the present tutorial to provide a comprehensive survey of the role of differential analysis
in systems and control (a partial account of which can be found in Section VI of \cite{Forni2014};
{see also the other paper of this tutorial session \cite{aminzare_sontag_tutorial_cdc2014}),
we will illustrate some questions that have stimulated a renewed interest for  differential analysis
in the recent years. The interested reader is also referred to the two-part invited session of CDC2013
for a sample of recent developments in that area.

Owing to the tutorial nature of the paper, the discussion will be exclusively restricted to
the  classical (adimensional) nonlinear pendulum model
\begin{equation}
\label{eq:pendulum}
\Sigma:\left\{
\begin{array}{rcl}
\dot{\vartheta} &=& v \\
\dot{v} &=& -\sin(\vartheta) - k v + u
\end{array}
\right. 
\quad (\vartheta,v) \in \calX := \mathbb{S}\times \real \ ,
\end{equation}
where $k\geq0$ is the damping coefficient and $u$ is the torque input. The specific
aim of the paper is therefore to understand as much as possible of the {\it global} behavior of model (\ref{eq:pendulum})
from its linearized dynamics ($(\delta \vartheta, \delta v) \in T_{(\vartheta,v)}\calX$)
\begin{equation}
\label{eq:closed_linearization}
\mymatrix{c}{ \dot{\delta \vartheta} \\ \dot{\delta v} } =
\underbrace{\mymatrix{cc}{ 0 & 1 \\ -\cos(\vartheta) & - k }}_{=:A(\vartheta,k)} 
\mymatrix{c}{ \delta \vartheta \\ \delta v } + \mymatrix{c}{ 0 \\ \delta u} 
\end{equation}
where any solution $(\delta \vartheta(\cdot),\delta v(\cdot))$ lives
at each time instant $t$ 
in the tangent space $T_{(\vartheta(t),v(t))}\calX$, where
$(\vartheta(\cdot),v(\cdot))$ is a solution to \eqref{fig:pendulum}.

The nonlinear pendulum model is an archetype example of nonlinear systems analysis.
As a control system, it is one of the simplest examples of nonlinear mechanical models
and many of its properties extend to more complex electro-mechanical models such as models
of robots, spacecrafts,  or electrical motors. As a dynamical system, it is one the simplest models to exhibit a rich and possibly complex
global behavior, owing to the interplay between small oscillations and large oscillations, two markedly
distinct behaviors for which everyone has a clear intuition developed since childhood. 

At the onset, it is worth observing that the pendulum is a nonlinear model for two related but distinct
reasons: the vector field is nonlinear due to the sinusoidal nature of the gravity torque but also the
state-space is nonlinear due to the angular nature of the pendulum position. In fact it could be argued
that the nonlinearity of the space is more fundamental than the nonlinearity of the vector field
in that example, and this feature of the pendulum extends to most nonlinear models encountered
in engineering.  The differential analysis, which linearizes both the space and the vector field, is 
perhaps especially relevant for such models.

\begin{figure}[htbp]
\centering
\includegraphics[width=0.6\columnwidth]{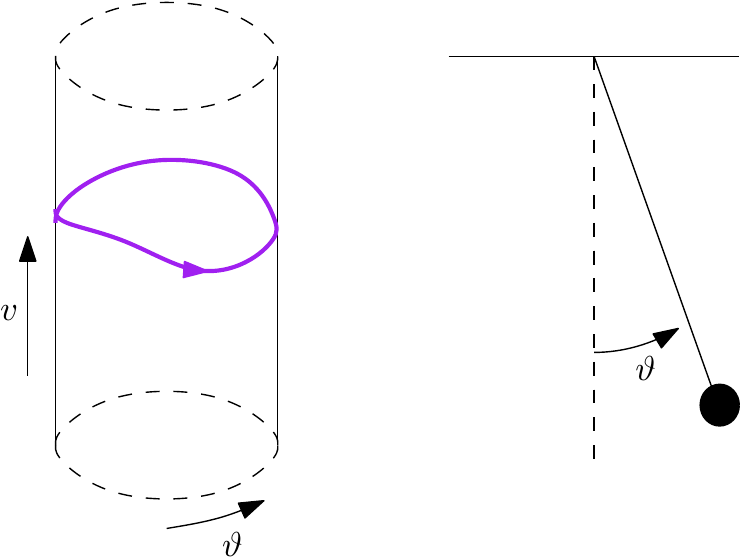}  
\caption{The natural state-space of the pendulum (left). The measure of the
angle $\vartheta$ (right).
}
\label{fig:pendulum}
\end{figure}

The  paper is organized as follows. Section II  
revisits the pendulum example from a classical nonlinear control perspective, 
pointing to  some limitations of nonlinear control that call for a differential
viewpoint. Section III revisits the pendulum example from a dynamical system perspective,
summarizing the main geometric properties of its limit sets.
Section IV introduces the differential analysis, starting with the classical role
of linearization in the local analysis of hyperbolic limit sets, and gradually
moving to the differential Lyapunov framework recently advocated by the authors \cite{Forni2014}.
The section concludes on a short discussion on horizontal contraction, a property
that purposely excludes specific directions in the tangent space from the  contraction analysis.
Section \ref{sec:positivity} illustrates on the pendulum example the novel concept of
differential positivity \cite{Forni2014a_ver1}, which is a projective form of differential contraction owing
to the positivity of the linearized dynamics. We will illustrate how differential positivity provides
a novel  tool for the differential analysis of limit cycles and, more generally, of one-dimensional attractors.

\section{A nonlinear control perspective}
\label{sec:classical}

\subsection{Feedback linearization and incremental dynamics}

The pendulum is feedback linearizable: the control input
 \begin{equation}
 \label{eq:feedback_linearization_input}
 u = \sin(\vartheta) + w
\end{equation}
transforms the nonlinear pendulum model into  the linear system
\begin{equation}
\label{eq:feedback_linearization}
\begin{array}{rcl}
\dot{\vartheta} &=& v \\
\dot{v} &=&  - k v + w \ .
\end{array} 
\end{equation}
Achieving linearity by feedback has been a cornerstone of nonlinear control theory
and is a key property for a regulation theory of nonlinear systems  \cite{Isidori1995}.

Exploiting linearity, it is straightforward to see that solving tracking or regulation 
problems on \eqref{eq:feedback_linearization} become trivial tasks in comparison 
to the fully nonlinear case.
The combination of the nonlinear cancellation in 
\eqref{eq:feedback_linearization_input} with a linear stabilizing feedback and a feedforward injection
would guarantee the asymptotic tracking of 
any suitable reference trajectory in $(\vartheta^*(\cdot),v^*(\cdot))\in\real \to \calX$.

A key difference between the nonlinear pendulum dynamics and the linear dynamics 
(\ref{eq:feedback_linearization}) is in the  {\it incremental} property: if 
$(\vartheta_1(\cdot),v_1(\cdot))$  and $(\vartheta_2(\cdot),v_2(\cdot))$ are two solutions of
 \eqref{fig:pendulum} for two different inputs $u_1(\cdot)$ and $u_2(\cdot)$, the 
 increment $(\Delta \vartheta(\cdot), \Delta v(\cdot))= (\vartheta_1(\cdot) - \vartheta_2(\cdot) ,
 v_1(\cdot)- v_2(\cdot))$ satisfies
 \begin{equation}
\label{eq:incremental}
\Delta \Sigma:\left\{
\begin{array}{rcl}
\dot{\Delta \vartheta} &=& \Delta v \\
\dot{\Delta v} &=& -(\sin(\vartheta_1) - \sin (\vartheta_2)) - k \Delta v +  \Delta u
\end{array}
\right. 
\end{equation}
 whose right-hand side differs from the original one. 
 Even the definition of the angular error $\Delta \vartheta$
 calls for some caution because of the nonlinear nature of angular variables. 
 
 The basic observation that the dynamics and incremental dynamics are equivalent only for linear systems
 is a fundamental bottleneck of nonlinear systems theory. Regulation, tracking, and observer design all
 involve the stabilization of the incremental dynamics. Only in linear system theory is the error between
 two arbitrary solutions equivalent to the error between one solution and the zero equilibrium solution.
 
 Feedback linearization makes the dynamics and the incremental dynamics equivalent. But if the 
 compensation of nonlinear terms by feedback is not possible, regulation theory becomes challenging
 even for the nonlinear pendulum, and requires {\it incremental} stability properties. This is a main motivation 
 for contraction theory, which seeks to exploit the stability properties of the linearized dynamics, that is, 
 the incremental dynamics for {\it infinitesimal} differences, in order to infer incremental stability properties.

\subsection{Energy-based Lyapunov control}
\label{sec:passivity}
Inherited from classical methods from physics, 
methods based on the conservation/dissipation of energy
are central in nonlinear control. The undamped pendulum
preserves the sum of kinetic and potential energy
\begin{equation}
\label{eq:energy}
E := \frac{v^2}{2} + \cos(\vartheta)
\end{equation}
during its motion, while it dissipates energy when the damping is nonzero $k > 0$. 
For open systems, dissipativity theory relates the energy dissipation to an  
external power supply \cite{Willems1972,Willems1972a}:
the energy is an internal {\it storage} that satisfies the balance 
\begin{equation}
\label{eq:dissip}
\dot E \leq  u \dot{\vartheta}  
\end{equation}
meaning that its rate of growth cannot exceed the mechanical power {\it supplied} to the system. 
Using $y:= \dot{\vartheta}$ to denote the output of the system, dissipativity with the 
supply $u y $ is a {\it passivity} property. Passivity is closely related to Lyapunov stability. The static output
feedback $u=-y$ adds {\it damping} in the system and is often sufficient to achieve asymptotic stability of the 
minimum energy equilibrium. For open systems, the supply rate
measures the effect of exogenous signals on the internal energy of the system.

Passivity based control is a building block of nonlinear control theory and has led to far reaching
generalizations in the theory of port-Hamiltonian systems \cite{Schaft2006,Ortega2001}, leading to an
interconnection theory for the energy-based stabilization of electro-mechanical systems. For instance,
the fundamental interconnection property that the feedback interconnection of passive systems
provides a direct solution to the PI control of passive systems because a PI controller is a passive system.

But a bottleneck of passivity theory is the generalization from stabilization to tracking control.
Fundamentally, this is because the dissipativity relationship seems of no direct use to analyze
the stability properties of the incremental dynamics. The energy -- or the storage -- provides
a natural distance between an arbitrary state and the state of minimum energy but it does not
provide a natural distance between two arbitrary solutions.

\subsection{Lure systems and Kalman conjecture}
Figure \ref{fig:euclidean_pendulum} illustrates that the nonlinear pendulum is a Lure system, that is,
it admits the feedback representation of a linear system with a static nonlinearity. The analysis
of Lure systems is another building block of nonlinear system theory, allowing to exploit the frequency-domain
properties of   the linear system in  the stability analysis of the nonlinear system.  

\begin{figure}[htbp]
\centering
\includegraphics[width=0.5\columnwidth]{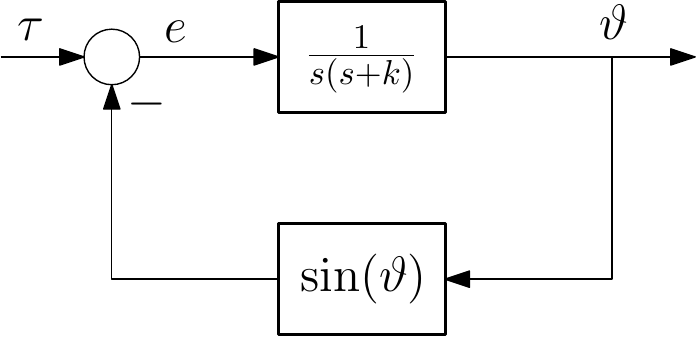}  
\caption{The representation of the pendulum in $\real^2$ as the negative feedback loop of a 
linear system with a sector nonlinearity.
}
\label{fig:euclidean_pendulum}
\end{figure}

Absolute stability theory seeks to characterize sufficient conditions of the static nonlinearity
to guarantee stability of the feedback system.  Most conditions for absolute stability would not
apply to the pendulum because they consider a static nonlinearity in a linear space, whereas
the sinusoidal nonlinearity should be considered as a static map defined on the circle.

But one relevant exception is the work of Kalman, which formulates conditions
on   the linearization of the nonlinearity. For a static nonlinearity satisfying
the condition $a \leq \sigma'(y) \leq b$, Kalman conjectured  stability of
the nonlinear system  if the feedback system
is stable for any constant gain $k \in [a,b]$, \cite{Kalman1957}.

Kalman's conjecture is a particular case of the Markus-Yamabe conjecture
\cite{Markus1960,Chamberland1997}, which infers
global asymptotic stability properties for the nonlinear system $\dot x = f(x)$ from
stability of the ``pointwise''  linearization $\dot{\delta x} = \partial f(x) \delta x$ at any point $x$.
A counter-example to Kalman conjecture eventually disproved both conjectures \cite{Leonov2010}
but the attempt is a typical example of differential analysis: global properties
of the nonlinear system are inferred from local analysis of the infinitesimal properties.

\section{A dynamical systems perspective}
\label{sec:dynamical_systems}

\subsection{Limits sets and bifurcations}

For a fixed constant torque input, the pendulum model is a two-dimensional
system that can be studied using phase portrait techniques, allowing for
a complete characterization of its limit sets.

Figure  \ref{fig:critical_damping} from \cite[Section 8.5]{Strogatz1994} summarizes
the possible asymptotic behaviors of the model as a function of two parameters:
 the damping coefficient $k$
and  the constant level of the torque $u$.
For large values of the damping, a constant input torque $ |u| \leq 1$ forces
the trajectories to converge to a fixed point: either the stable downward equilibrium,
or the unstable upward equilibrium for solutions initialized on the one-dimensional
stable manifold of this saddle point.  For a torque magnitude $|u| >1$,  the unique limit set is a globally attractive limit cycle.
For small damping $k$, solutions still converge either to a fixed point for small torque or to a limit cycle
for large torque, but  an intermediate region exists in the parameter space where the
stable limit cycle behavior coexists with the stable fixed point. This bistable behavior exists if the damping
parameter does not exceed a critical damping $k_c$.
The overall behavior is summarized in Figure \ref{fig:critical_damping}.
\begin{figure}[htbp]
\centering
\includegraphics[width=0.98\columnwidth]{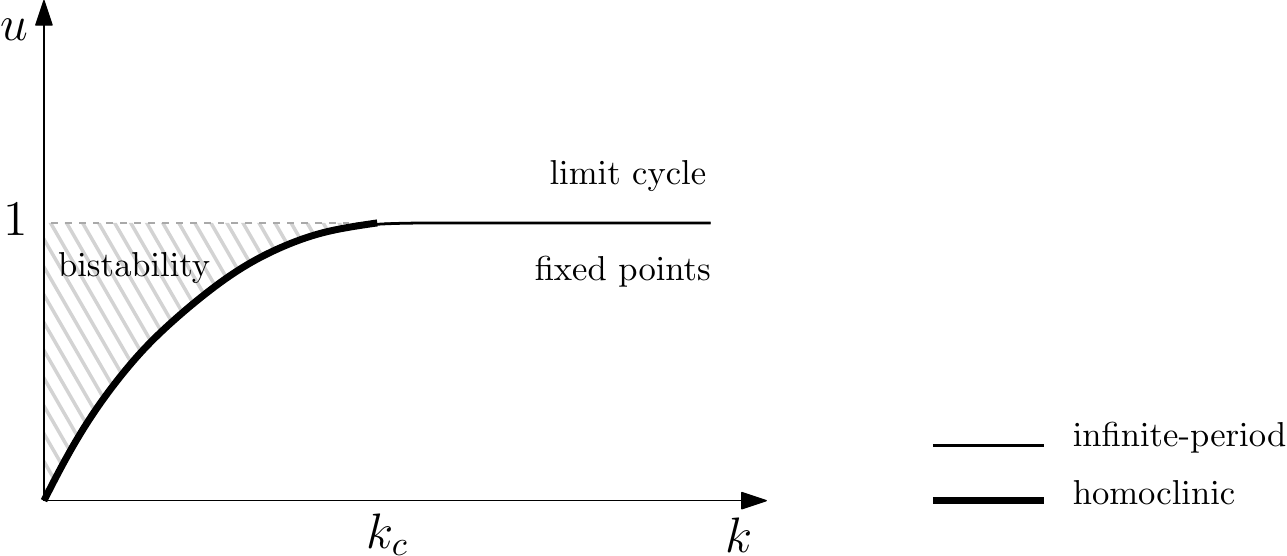}
\caption{The qualitative behavior of the pendulum for large/small values of torque and damping, as reported in \cite{Strogatz1994}}
\label{fig:critical_damping}
\end{figure}

Specific bifurcations delineate the different types of asymptotic behavior in the
parameter space. For large damping, the two fixed points existing for $u < 1$
approach each other as the torque is increased to eventually merge in a single
fixed point for $u=1$ in a  so-called
 infinite-period bifurcation \cite[Section 8.4]{Strogatz1994}.

A different bifurcation scenario gives rise to the bistable region in
Figure \ref{fig:critical_damping}. For any $k< k_c$ there exists a critical
value $u = u_c(k)$ for which the pendulum encounters
a \emph{homoclinic bifurcation}.
(see \cite[Section 8.5]{Strogatz1994} and
Figures 8.4.3, 8.5.7 and 8.5.8 therein).
For decreasing value of $u_c(k) < u < 1$ the limit cycle gets closer to
the unstable manifold of the saddle. 
At $u=u_c(k)$ the  limit cycle merges with the unstable manifold of
the saddle, which also coincides with the stable manifold of the saddle
(see Figure \ref{fig:homoclinic_simple}), and disappear for $u<u_c(k)$.
For $u_c(k) < u < 1$,
the stable manifold of the saddle is an important geometric object: it
separates the basin of attraction of the stable fixed point from the basin of attraction of the limit cycle.
\begin{figure}[htbp]
\centering
\includegraphics[width=0.6\columnwidth]{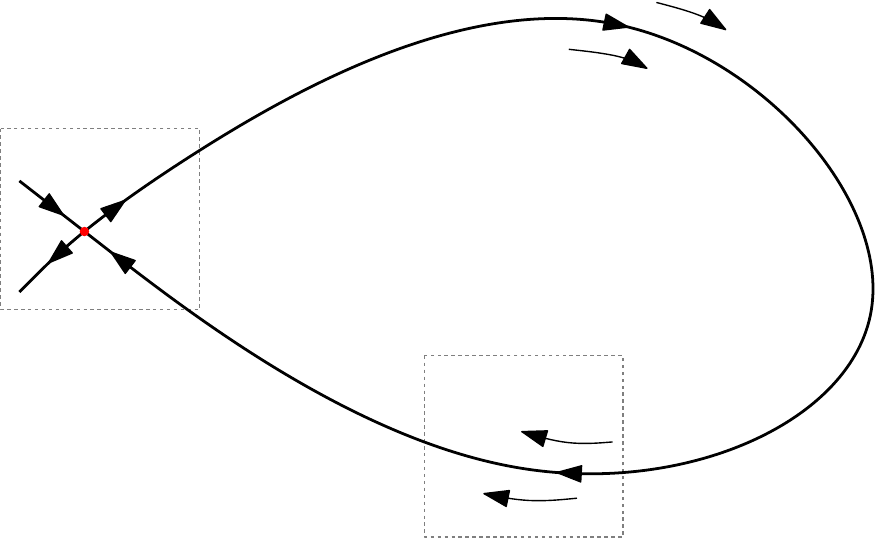}
\caption{A homoclinic horbit connects the unstable manifold of the saddle with its stable manifold.
The saddle point is the $\alpha$ and $\omega$ limit set of each point within the homoclinic orbit.}
\label{fig:homoclinic_simple}
\end{figure}

\subsection{The overdamped pendulum}
For large damping, further insight on the qualitative dynamics is provided by
singular perturbation analysis, which exploits the analysis of the singular
behavior obtained in the limit of infinitely separated time-scales.

Time scale separation on the pendulum is dictated by the damping coefficient.
In the overdamped limit, the two-dimensional behavior decouples into two one-dimensional
behaviors, which is a drastic simplification.
Following \cite[Section 4.4]{Strogatz1994}, in the overdamped limit $k\to \infty$,
the pendulum dynamics reduces to a first order (open gradient) dynamics represented by
the (normalized) equation
\begin{equation}
\label{eq:overdamped_pendulum}
 \dot{\vartheta} = -\sin(\vartheta) + u \ .
\end{equation}
For instance, in the overdamped limit the velocity component of \eqref{eq:pendulum}
reads $kv = -\sin(\vartheta)+u$. Thus, $k\dot{\vartheta} = -\sin(\vartheta) + u$
which, by time reparameterization $k\tau = t$, gives \eqref{eq:overdamped_pendulum}.
\footnote{For simplicity, in \eqref{eq:overdamped_pendulum}
we are denoting by $\dot{\vartheta}$ the quantity $\frac{d\vartheta}{d \tau}$.}

The one-dimensional model (\ref{eq:overdamped_pendulum}) captures the qualitative
behavior of the pendulum for large damping: it has two fixed points for $| u | <1$
and no fixed points, meaning a periodic behavior, for $| u | >1$. The saddle-node bifurcation at $u=1$
is the one-dimensional analog of the infinite period bifurcation of the two-dimensional pendulum.
In contrast, the bistable behavior of the pendulum for smaller damping is not captured under
the time-scale separation assumption.

\subsection{Ingredients for complex attractors}

The nonlinear pendulum is especially valuable as a prototype
example of dynamical systems textbooks in that it illustrates
a fundamental route to hyperbolic strange attractors: the
simple bistable behavior reviewed in the previous section for
small damping can be turned into a complex chaotic behavior
under a weak harmonic input of the type $u =  \epsilon \sin (\omega t)$.

This is because the saddle homoclinic orbit that exists in the range
of small damping and small torque: it
allows for {\it recurrence} of the saddle point neighborhood, that is, trajectories
that start close to the saddle point can return to the saddle point after a large excursion,
together with {\it sensitivity} of the initial condition: a small perturbation near the saddle point can
change the small or large oscillation fate of the trajectory.
Such behavior is the essence of Smale's construction of hyperbolic strange
attractors \cite[pp. 843-852]{Smale2000b} and the stable manifold theorem.

In that sense, the homoclinic orbit illustrated in Figure \ref{fig:homoclinic_simple}
is a fundamental ingredient of complex behaviors. And it has a particularly simple
and concrete interpretation in the nonlinear pendulum model
 as the geometric object that separates small oscillations from large oscillations
for small damping and small torque. The next sections will illustrate how this global property
can be captured in a differential framework.

\section{A differential perspective}
\label{sec:contraction}

\subsection{Linearization and local analysis}
Differential methods recognize that the analysis of the linearization of the system dynamics
along trajectories captures important properties of the system behavior. They are the essence
of  local stability analysis. The simplest case is provided
by Lyapunov's first method, for the analysis of the local stability properties of fixed points.
For the pendulum with zero torque, the linearization of the dynamics is given by
\eqref{eq:closed_linearization} and $\delta u = 0$.
The eigenvalues of the state matrix 
\begin{equation}
\label{eq:state_matrix_linearization}
A(\vartheta,k) = \mymatrix{cc}{ 0 & 1 \\ -\cos(\vartheta) & - k }
\end{equation}
at the fixed points
$(\vartheta,v) \in \{(0,0),(\pi,0)\}$ reveal that the fixed point in zero is
locally asymptotically stable, while the other fixed point is a saddle.

Lyapunov's first method rests on the observation that any trajectory $(\delta \vartheta(\cdot), \delta v(\cdot))$
of \eqref{eq:closed_linearization} at the fixed point $(\vartheta,v) = (0,0)$
is an approximation of the infinitesimal mismatch between the
trajectory $(\vartheta(t),v(t)) = (0,0)$ at equilibrium and the trajectory
$(\hat{\vartheta}(\cdot),\hat{v}(\cdot))$ arising from an infinitesimal initial variation
given by $\hat{\vartheta}(t_0) - \vartheta(t_0) = \delta \vartheta(t_0)$ and
$\hat{v}(t_0) - v(t_0) = \delta v(t_0)$. Indeed, exponential stability of the linearization
implies asymptotic convergence of $(\hat{\vartheta}(\cdot),\hat{v}(\cdot))$ to
the fixed point. In that sense, the linearization captures the infinitesimal incremental
dynamics in the neighborhood of a particular solution.

A similar approach captures the local stability properties of limit cycles.
Let $(\vartheta(\cdot),v(\cdot))$ be the periodic trajectory of the pendulum
for some $u>1$. Periodicity reads:
there exists a time interval $T>0$ such that
$(\vartheta(t+T),v(t+T)) = (\vartheta(t),v(t))$ for all $t$.
The fundamental matrix solution $\Phi(\cdot)$ of the linearization \eqref{eq:closed_linearization}
along the periodic trajectory $(\vartheta(\cdot),v(\cdot))$ satisfies
\begin{equation}
\label{eq:floquet1}
\dot{\Phi}(t) = A(\vartheta(t),k) \Phi(t)
\end{equation}
that, by periodicity, leads to the identity $\Phi(t+T) =  \Phi(t)$ \ .
Considering the initial condition $\Phi(0) = I$ ($I$ is the identity matrix),
the eigenvalues $\rho_1, \dots, \rho_n$ of the update map
\begin{equation}
\Delta_T := \Phi(T)
\end{equation}
are the 
\emph{characteristic Floquet multipliers} of 
the periodic trajectory $(\vartheta(\cdot),v(\cdot))$. 
These eigenvalues characterize the 
behavior of the nonlinear pendulum in the neighborhood of the
periodic trajectory \cite[Section 1.5]{Holmes1983}.

Looking at Figure \ref{fig:poincare},
in an infinitesimal neighborhood of the periodic trajectory, the update
map captures the convergence among neighboring trajectories
crossing the Poincar{\'e} section transversal to the system flows.
Indeed, $n-1$ Floquet multipliers smaller than one
imply local asymptotic stability of the limit cycle (by symmetry,
one multiplier is necessarily equal to one).
\begin{figure}[htbp]
\centering
\includegraphics[width=0.6\columnwidth]{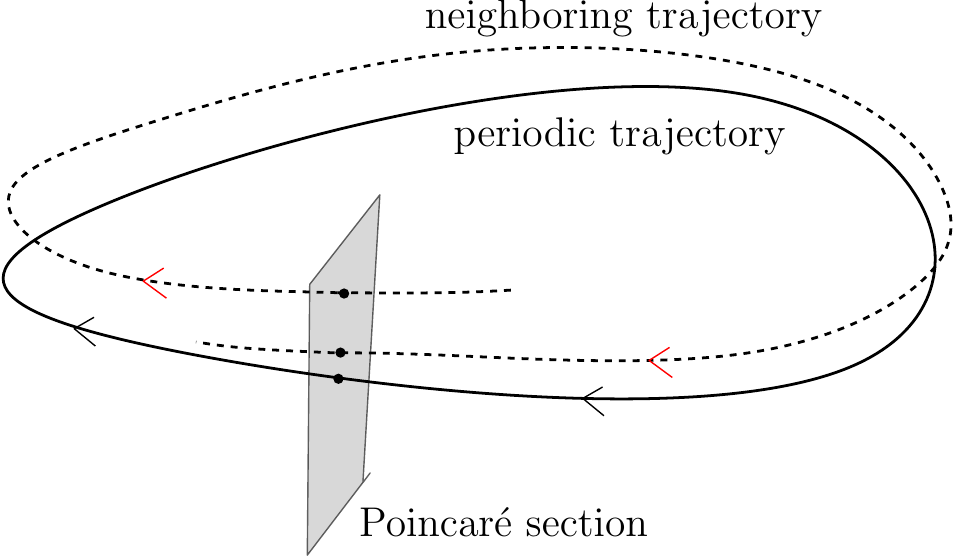}
\caption{$n-1$ Floquet multipliers smaller than one guarantees
contraction of the Poincar{\'e} map, thus local attractiveness of the limit cycle.}
\label{fig:poincare}
\end{figure}

The study of the linearized dynamics plays a fundamental role
also in the characterization of chaotic behaviors, through
the notion of Lyapunov exponents, \cite{Arnold1986,Lyapunov_exponent}.
The maximal Lyapunov exponent is a measure of the
maximal separation rate between
two infinitesimally close trajectories,
which makes contact with the
sensitivity of trajectories with respect to initial conditions.
The maximal Lyapunov exponent is
captured by the growth rate of the fundamental solution,
which for the pendulum reads
\begin{equation}
\label{eq:lyapunov_exponent}
\lim_{t\to\infty} \frac{1}{t} \ln |\Phi(t)|
\end{equation}
computed along any system trajectory $(\vartheta(\cdot),v(\cdot))$
from the initial condition $\Phi(0)=I$.
Clearly, the maximal Lyapunov exponent depends on the
particular trajectory $(\vartheta(\cdot),v(\cdot))$ along which the linearization is computed.
The limit in \eqref{eq:lyapunov_exponent} clarifies, however, that such a dependence
is related to the particular attractor to which the trajectory converge.
The selection of a particular matrix norm may change the value of the
maximal Lyapunov exponent. For systems with bounded trajectories, a positive
maximal Lyapunov exponent is an indicator of possible chaotic behaviors.

\subsection{From Kalman's conjecture to differential Lyapunov theory}

The aim of differential analysis is to
exploit the properties of the  linearized dynamics beyond the local
stability analysis of attractors. Kalman's conjecture and Markus-Yamabe conjecture
illustrate attempts  to infer global properties of the nonlinear system from
the analysis of linearized dynamics.

The conditions of the Kalman's conjecture are based
on the linearized dynamics of a Lure system
$\dot x = f(x):= A x - B \sigma (C x) $ where $(A,B,C) $ is a
minimal state-space representation of the linear system
in feedback interconnection with the static nonlinearity $\sigma(\cdot)$.
Requiring that the matrix $(A-\mu BC)$ is Hurwitz
for any $\mu$ is equivalent to check that the Jacobian matrix
$\partial f(x)$ is Hurwitz for any $x$, showing that Kalman's conjecture is a particular
case of the Markus-Yamabe conjecture \cite{Kalman1957,Markus1960,Chamberland1997}.   
It is well known that a Huwitz Jacobian matrix does not
guarantee stability \cite[Perron Effects]{Leonov2008a}, \cite{Leonov2007}. 
In fact, within the
reformulation based on the Jacobian, Kalman's condition reads 
\begin{equation}
\label{eq:lyapunov}
\partial f(x)^T P(x) + P(x) \partial f(x) < 0 \qquad \forall x \ ,
\end{equation}
where $P(x)$ is a positive and symmetric matrix for each $x$.
However, the variation of $P(x)$ cannot be neglected, which
requires the satisfaction of the extended condition
\begin{equation}
\label{eq:riemann}
\partial f(x)^T P(x) + P(x) \partial f(x) + \dot{P}(x)< 0 \qquad \forall x \ ,
\end{equation}
where $\dot{P}(x)$ represents the variation of the ``metric'' along
the vector field $f(x)$.

The gap between the Jacobian conjecture and a sufficient condition for global
asymptotic stability thus relates to analyzing the stability properties of the {\it frozen}
linearized dynamics at every point instead of along a specific trajectory.

The ``stability'' condition \eqref{eq:riemann} allows for an interesting geometric reinterpretation
when the matrix $P(x)$ is the representation, in coordinates, of a Riemannian tensor.
\eqref{eq:riemann} provides a coordinate formulation of the contraction of the
Riemannian tensor along the flow of the system. As a consequence,
given a positive and symmetric matrix $P(x)$, smooth in $x$,
condition \eqref{eq:riemann}
guarantees not only that the fixed point of the nonlinear dynamics is asymptotically
stable, but also that the nonlinear dynamics is  contractive, that is, any pair of solutions
$x(\cdot), z(\cdot)$ of the differential equation $\dot{x} = f(x)$ satisfy
\begin{equation}
\label{eq:contraction}
 \lim_{t\to\infty} d(x(t),z(t)) = 0 \ ,
\end{equation}
where $d$ is the Riemannian distance provided by the Riemannian tensor by integration along geodesics.
Interestingly, if the induced distance and the state space of the system define a complete
metric space, the existence of a stable fixed point follows from the contraction mapping theorem.
This is the essence of the {\it contraction analysis} advocated by Lohmiller and Slotine, \cite{Lohmiller1998}.

The key observation is that the \emph{asymptotic stability} of the linearized dynamics
guarantees that the nonlinear system is \emph{contractive}. It follows that
the nonlinear dynamics may have at most one fixed point which is a global attractor for the system dynamics.
Looking at Figure \ref{fig:contraction2}, the intuition is that the motion of neighboring trajectories is
described by the linearized dynamics, and their convergence is captured by the asymptotic stability property
of the linearized dynamics. By patching many neighboring trajectories, i.e. by integration 
along differentiable curves connecting different trajectories, 
the local contraction among
neighboring trajectories translates into contraction among any pair of trajectories.
\begin{figure}[htbp]
\centering
\includegraphics[width=0.7\columnwidth]{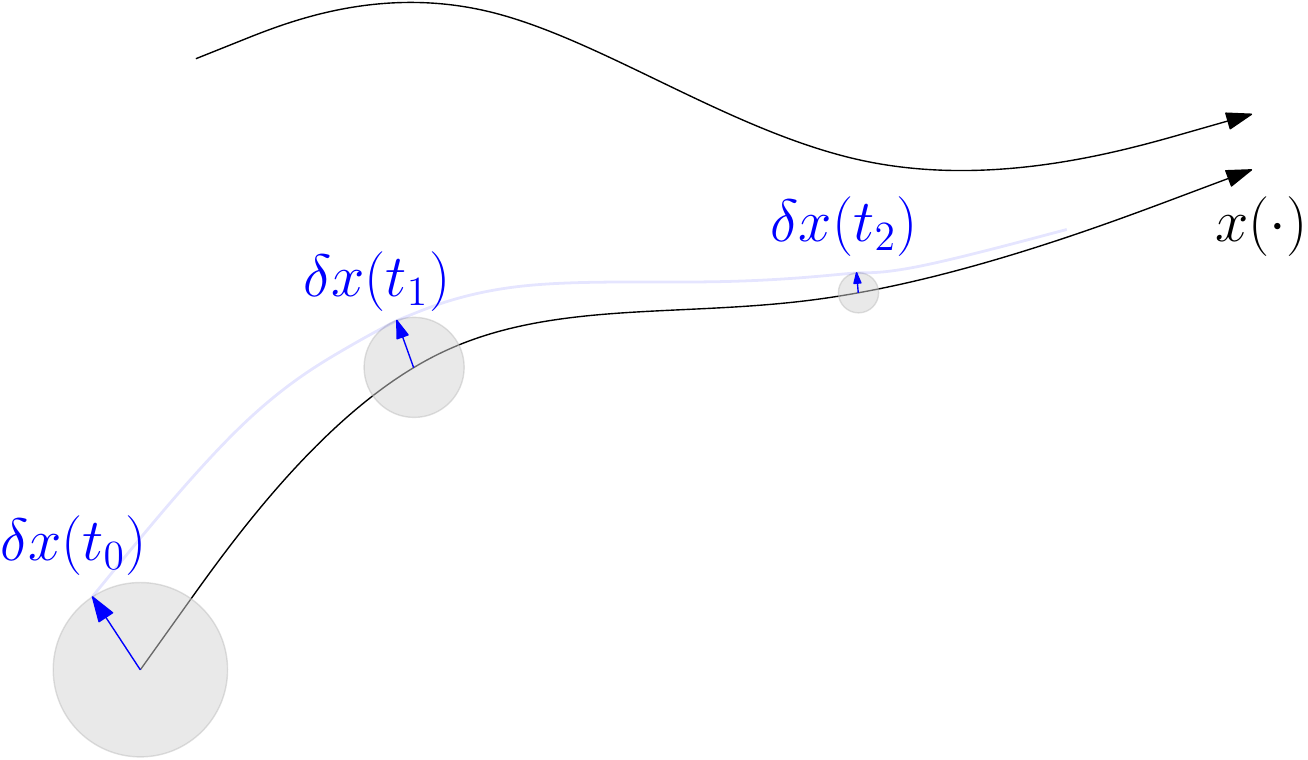}
\caption{The convergence of infinitesimal neighboring trajectories towards each other is captured by
the asymptotic stability of the linearized dynamics.}
\label{fig:contraction2}
\end{figure}

From a control-theoretic perspective, the condition \eqref{eq:riemann} based on Riemannian metrics
share the structure of classical Lyapunov stability with respect to quadratic Lyapunov functions.
The step forward recently proposed in \cite{Forni2014} is to view contraction analysis as a differential
Lyapunov theory, allowing to consider more general (and not necessarily quadratic) Lyapunov functions
in the tangent bundle. 

Let $\calX$ be the state-space of the system  represented by 
$\dot{x} = f(x)$, and consider the \emph{prolonged system} \cite{Crouch1987}
represented by the pairing of the system dynamics with the linearized dynamics 
\begin{equation}
 \left\{
 \begin{array}{rcl}
 \dot{x} &=& f(x) \\
 \dot{\delta x} &=& \partial f(x) \delta x \\
 \end{array}
 \right. \qquad (x,\delta x) \in T\calX \ .
\end{equation}
$T\calX$ denotes the tangent bundle of $\calX$.
In analogy with classical Lyapunov theory,
a Finsler-Lyapunov function from the tangent bundle $T\calX$ to $\real_{\geq 0}$ satisfies the bounds
\begin{equation}
\label{eq:finslyap}
 c_1 |\delta x|_x^p \leq V(x,\delta x) \leq c_2 |\delta x|_x^p 
\end{equation}
where $c_1, c_2 \in \real_{>0}$, $p$ is some positive integer and $|\cdot|_x$ is a Finsler metric.
Intuitively, $|\cdot|_x$ defines a Minkowski norm
in each tangent space $T_x\calX$, \cite{Shen00} 
\footnote{On vector spaces $\calX = \realn$, 
the tangent bundle can be identified with $T\calX = \realn\times\realn$ and 
any norm $|\cdot|$ in $\realn$ provides a \emph{constant} Finsler metric $|\cdot|_x := |\cdot |$.}
From \eqref{eq:finslyap},
a Finsler-Lyapunov function measures the length of any tangent vector $\delta x$. In other words, $V$ is a measure
of the distance of $\delta x$ from $0$, providing to the linearized dynamics the equivalent
of a classical Lyapunov function, typically measuring the distance of the state $x$
from the $0$ equilibrium.

The stability of the linearized dynamics along the system trajectories follows from the pointwise
decay of the Finsler-Lyapunov function along the trajectories of the prolonged system.
Geometrically,
one has to establish
\begin{equation}
\label{eq:finslyap_decay}
\dot{V} \leq - \alpha(V(x,\delta x)) 
\end{equation}
where $\dot{V}$ reads
$\partial_x V(x,\delta x) f(x) + \partial_{\delta x} V(x,\delta x) \partial f(x) \delta x$
and $\alpha$ is a $\calK$ function.
\eqref{eq:finslyap} and \eqref{eq:finslyap_decay}
guarantee that the nonlinear system is  contractive \eqref{eq:contraction} 
but
with respect to the Finslerian distance $d$ induced by the Finsler metric $|\cdot|_x$,
by integration, \cite[Theorem 1]{Forni2014}. A straightforward corollary is that any fixed point
of the nonlinear system is necessarily unique and globally asymptotically stable.

\eqref{eq:finslyap} and \eqref{eq:finslyap_decay} subsume many conditions
for contraction available in the literature, \cite{Lewis1949,Lohmiller1998,Angeli00,Rouchon2003,Pavlov04,Jouffroy05,Russo10,SimpsonPorco2014}.
For a detailed comparison, please refer to \cite[Section VI]{Forni2014}.
See also \cite{aminzare_sontag_tutorial_cdc2014} for a discussion on
the basic concepts of contraction theory.

\eqref{eq:finslyap} and \eqref{eq:finslyap_decay} allows for the analysis of 
time-varying systems $\dot{x} = f(x,t)$, for which $\dot{V}$ reads
$\partial_x V(x,\delta x) f(x,t) + \partial_{\delta x} V(x,\delta x) \partial_x f(x,t) \delta x$.
The notion of Finsler-Lyapunov function in  \eqref{eq:finslyap} can be further generalized to time-varying functions $V$,
modifying accordingly \eqref{eq:finslyap_decay}. 
By exploiting the analogy with classical Lyapunov theory, under boundedness assumption on
the trajectories of the (time-invariant) nonlinear system $\dot{x} = f(x)$, 
\eqref{eq:finslyap_decay} can be relaxed to the LaSalle-like formulation
\begin{equation}
\label{eq:LaSalle}
\dot{V} \leq - \alpha(x,\delta x) 
\end{equation}
where $\alpha:T\calX \to \real_{\geq 0}$. Then, the contraction property \eqref{eq:contraction}
holds provided that the largest invariant set contained in
\begin{equation}
\Pi(x,\delta x) := \{(x,\delta x)\in T\calX\,|\,  \alpha(x,\delta x)=0\}
\end{equation}
is given by $\calX \times \{0\}$, \cite[Theorem 2]{Forni2014}.

\subsection{Differential analysis of the overdamped pendulum}

The overdamped pendulum \eqref{eq:overdamped_pendulum} is studied in \cite{Forni2014}
via differential analysis. For $u=0$,
the simple choice of the Finsler-Lyapunov function $V:= \delta \vartheta^2$ guarantees that
\begin{equation}
\label{eq:decay1}
\dot{V} = -\cos(\vartheta) \delta \vartheta^2 < 0 \qquad \forall (\vartheta,\delta \vartheta) \in \left(-\frac{\pi}{2},\frac{\pi}{2}\right)\times \real.
\end{equation}
The decay of the Finsler-Lyapunov function is restricted to the open lower half of the circle. Thus,
the contraction \eqref{eq:contraction} holds only among those trajectories of the nonlinear dynamics whose
image is contained within $\left(-\frac{\pi}{2},\frac{\pi}{2}\right)$ (forward invariant region). The particular selection of a \emph{constant}
Finsler-Lyapunov function with respect to $\vartheta$ (in coordinates)
makes the condition $\dot{V} < 0$ feasible only within the region of strict monotonicity of the vector
field $\dot{\vartheta} = - \sin(\vartheta)$. For instance, this is the result that one would obtain by considering the
convergence to zero of the arc length $\Delta \vartheta := \vartheta_1 - \vartheta_2$, where $\vartheta_1,\vartheta_2$ both
satisfy the overdamped pendulum dynamics.

The Finsler-Lyapunov function $V:= \frac{\delta \vartheta^2}{1+\cos(\vartheta)}$ deforms the measure of the length of $\delta \vartheta$ as a function
of the particular point $\vartheta$, establishing contraction beyond monotonicity of the right-hand side of overdamped pendulum equations.
For this new Finsler-Lyapunov function \eqref{eq:finslyap_decay} reads
\begin{equation}
\label{eq:decay2}
\dot{V} = - \delta \vartheta^2 < 0 \qquad \forall (\vartheta,\delta \vartheta) \in \left(-\pi,\pi\right)\times \real,
\end{equation}
that is, for all the points of the circle but the unstable point at $\pi$. From \eqref{eq:contraction}, it is clear that the exclusion of the unstable fixed point
is a necessary condition to achieve the decay of the Finsler-Lyapunov function, since no
trajectory converges to the steady-state trajectory $\pi$. 
Looking at Figure \ref{fig:overdamped2}, the intuitive
explanation for \eqref{eq:decay2} is that 
the distance $d$ associated to the new Finsler-Lyapunov function by integration of
the Finsler metric $\left(\frac{\delta x^2}{1+\cos(\vartheta)}\right)^{\!\frac{1}{2}}$ measures
constant arc length intervals $b-a = c-b = \mathit{const}$ in a way that guarantees
$d(a,b) < d(b,c)$, following a transformation similar to the one represented in
Figure \ref{fig:overdamped2}.
\begin{figure}[htbp]
\centering
\includegraphics[width=0.7\columnwidth]{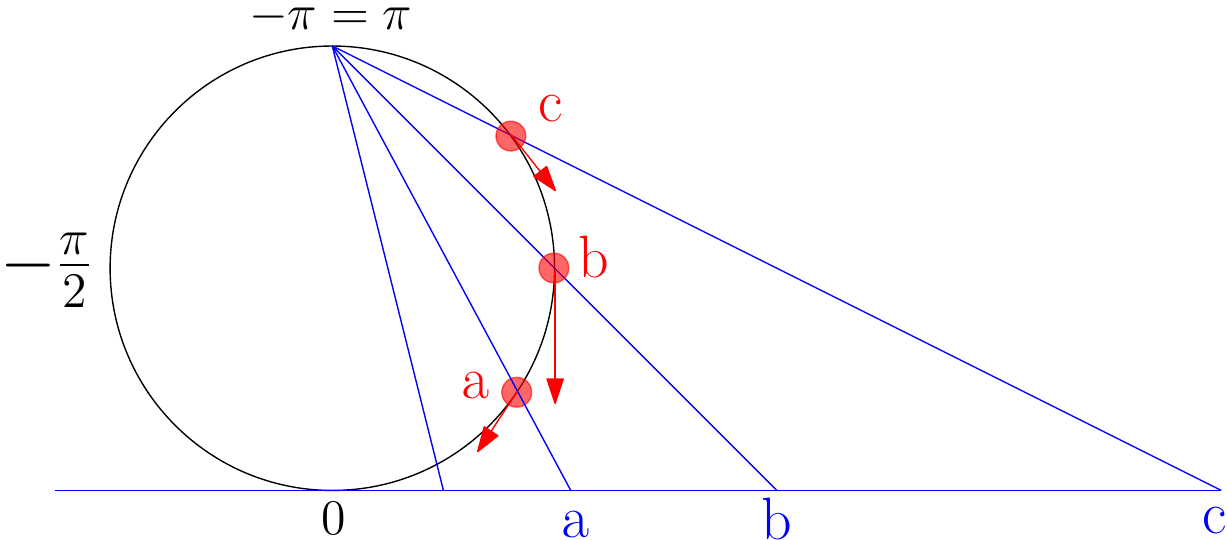}
\caption{A representation of the geodesic distance induced by the Finsler metric $\sqrt{\frac{\delta x^2}{1+\cos(\vartheta)}}$ on the
circle.}
\label{fig:overdamped2}
\end{figure}

It is insightful to look at the overdamped pendulum under the feedforward action of the 
(possibly non-constant) torque $u\neq 0$,
 to illustrate the use of Finsler-Lyapunov functions away from the study of stability of fixed points.
\eqref{eq:closed_linearization} with $\delta u = 0$ characterize the linearized dynamics along any trajectory 
$(\vartheta(\cdot),v(\cdot))$ generated by the action of the input $u(\cdot)$. 
In fact, for $\delta u = 0$, the linearization captures the infinitesimal 
mismatch between $(\vartheta(\cdot),v(\cdot))$ and any other neighboring trajectory generated by the same input $u(\cdot)$.
For the pendulum, the presence of the input changes the pointwise decay \eqref{eq:decay2}
into
\begin{equation}
\label{eq:decay3}
\dot{V} = - \delta \vartheta^2 + w(\vartheta,\delta \vartheta,u)  \qquad \forall (\vartheta,\delta \vartheta) \in \left(-\pi,\pi\right)\times \real,
\end{equation}
where the term $w$ is not sign definite. Indeed, not surprisingly, the trajectories along which the linearized dynamics
are now modified by the action of the input $u$, and the decay of a non-constant Finsler-Lyapunov function
designed by taking into account the specific nonlinearities of the system vector field  is perturbed by the action of  the input.
To achieve the property of \emph{uniform} asymptotic stability of the linearization - or \emph{uniform contraction} -
with respect to the input, the input action must be paired
to the particular definition of the Finsler-Lyapunov function. For example, for the overdamped pendulum, taking $u = \cos(\frac{\vartheta}{2})r$
guarantees that the inequality \eqref{eq:decay2}  holds uniformly in $r$, that is,
\begin{equation}
\label{eq:decay3b}
w(\vartheta,\delta \vartheta,\cos(\vartheta/2)r)= 0 
\end{equation}
for all $(\vartheta,\delta \vartheta) \in \left(-\pi,\pi\right)\times \real$ and all $r\in \real$,
as detailed in \cite[Example 1]{Forni2013a}.

The uniform contraction of the overdamped pendulum is illustrated by the simulations in Figure \ref{fig:entrainment}, for small and large
sinusoidal signals $r$. Uniform contraction with respect to the input is a powerful property, at the core of many results in contraction-based design \cite{Lohmiller1998,Pavlov2005,Pavlov2007,Pavlov2008,Russo10,Sontag2010,Forni2013a}.
A uniform contracting system behaves like a filter: it forgets the initial conditions and its trajectories asymptotically converge to the
unique, globally attractive steady state compatible with the input signal, for any given input signal injected into the system. 

\begin{figure}[htbp]
\centering
\includegraphics[width=0.485\columnwidth]{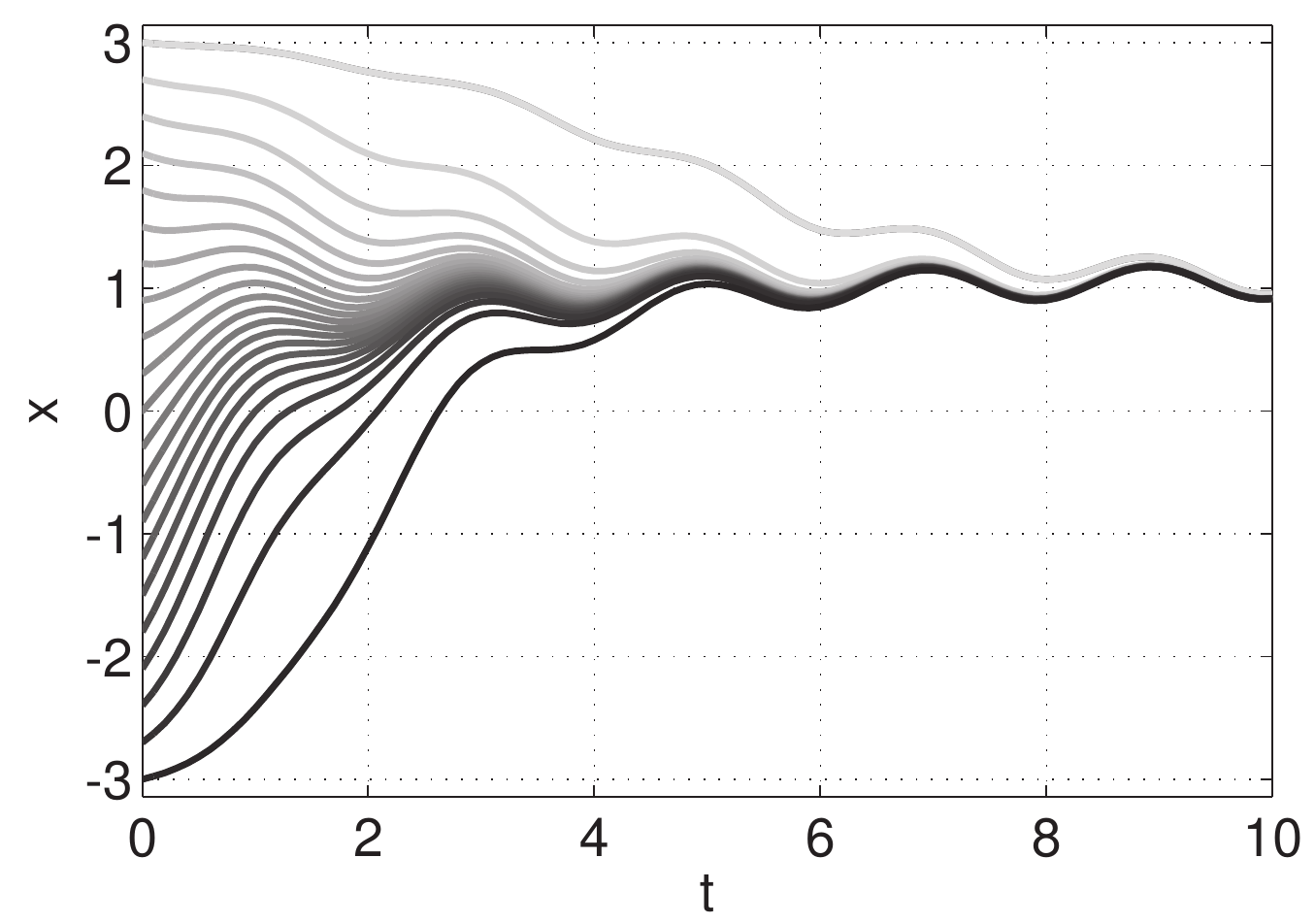}
\includegraphics[width=0.485\columnwidth]{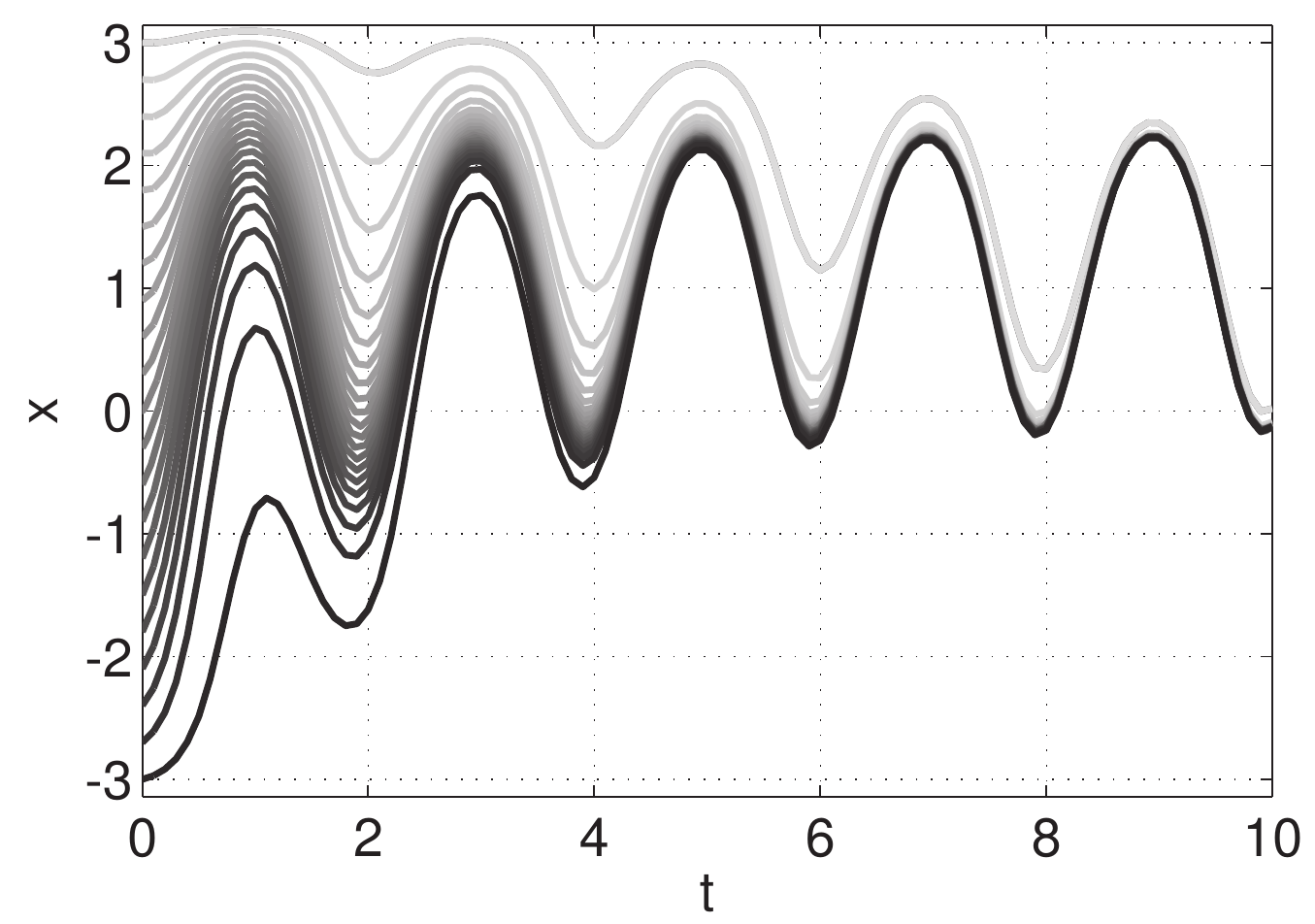}
\caption{Entrainment of the overdamped pendulum for $u = \cos(\frac{\vartheta}{2}) r$ and 
oscillating exogenous signal
$r = 1 + \gamma\sin(\pi t)$, for constant gain $\gamma$ small (left) and large (right).}
\label{fig:entrainment}
\end{figure}

Uniform contraction is also at the root of several contributions on the interconnection of contractive nonlinear systems.
As in classical control, the system arising from the interconnection of  contractive systems is not necessarily  contractive.
The results available in the literature extend to the differential framework classical 
cascade and small gain approaches  \cite{Sontag2010,Russo2013,SimpsonPorco2014} and 
dissipativity theory \cite{Schaft2013,Forni2013,Forni2013a}. 
For example, without entering into the details of the analysis (the reader is referred to \cite[Example 1]{Forni2013a}), 
the overdamped pendulum 
with $u= \cos(\frac{\vartheta}{2})r$ is \emph{differentially passive} from $r$ to the (differentially) passivating
output $y := \int_0^\vartheta \sec(\frac{s}{2}) d s $, that is,
\begin{equation}
 \dot{V} \leq \delta r \delta y \ .
\end{equation}
In analogy with classical passivity in Section \ref{sec:passivity}, $V$ has the role here of 
\emph{differential storage} whose variation is bounded by the \emph{differential supply} $\delta r \delta y$.

The analogy with classical passivity goes beyond basic definitions: the negative feedback interconnection of differentially 
passive systems is differentially passive. Thus, for example, the closed loop of the overdamped pendulum 
with any strictly increasing static nonlinearity $r = -h(y)$, that is, $\delta r = \partial h(y)\delta y$ with $\partial h(y) > 0$,
leads to a contractive dynamics. Figure \ref{fig:passivity} illustrates the behavior of the contractive system arising from
the interconnection of two overdamped pendulums.

\begin{figure}[htbp]
\centering
\includegraphics[width=0.48\columnwidth]{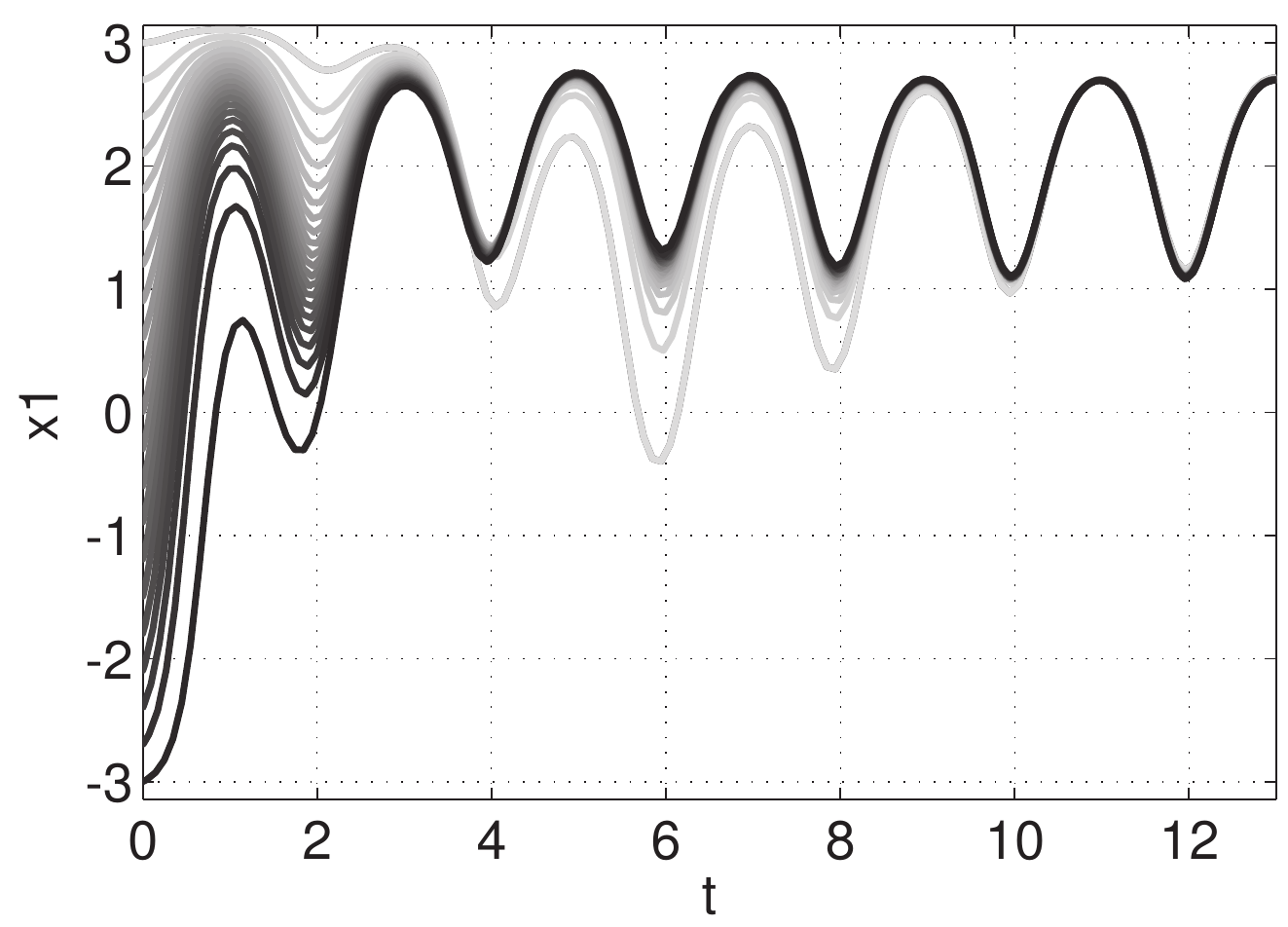}
\includegraphics[width=0.48\columnwidth]{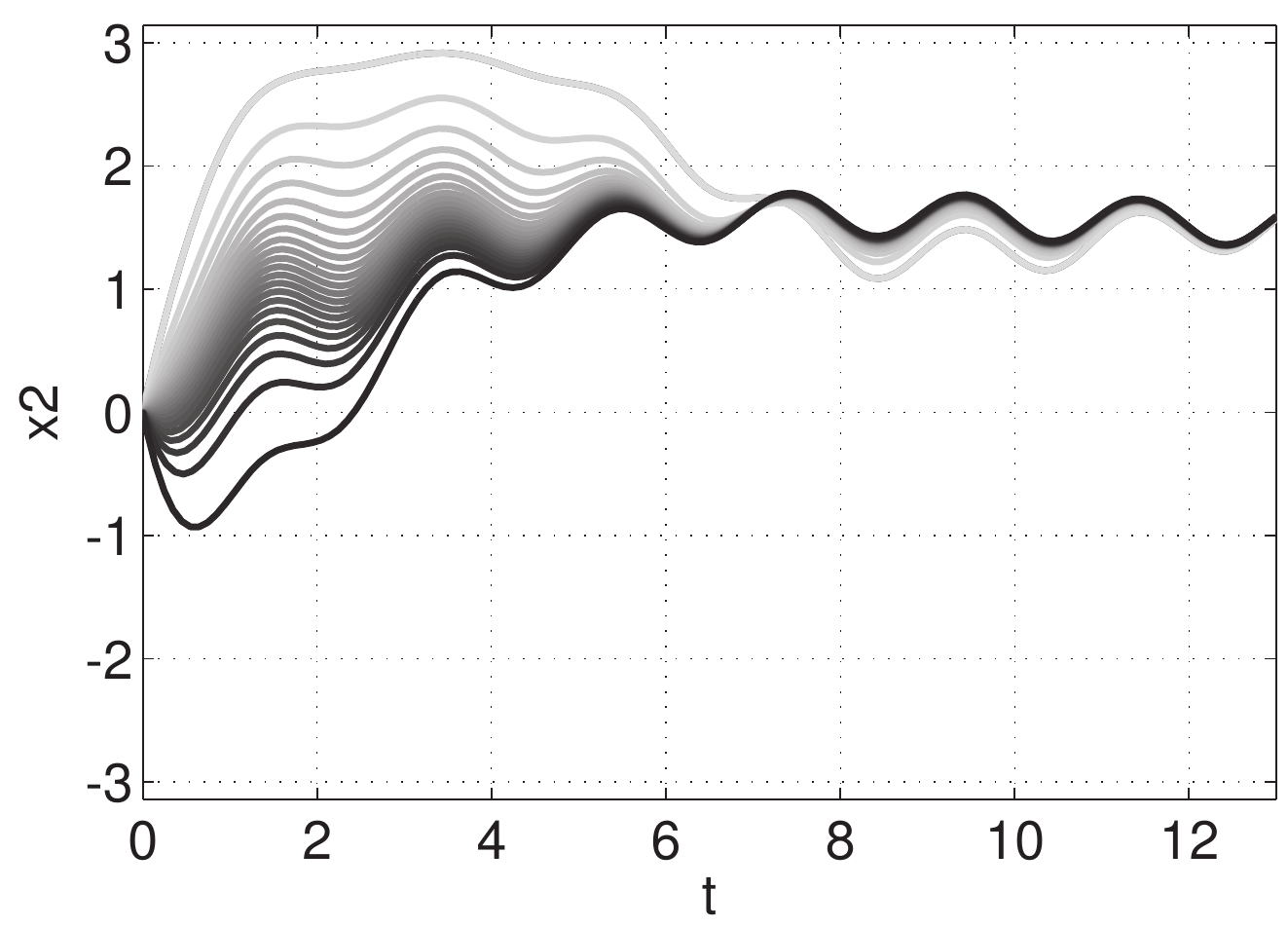}
\caption{Feedback interconnection $r_1 = -y_2 + q_1$, $r_2 = y_1+q_2$
of two differentially passive overdamped pendulums, where the indices $1$ and $2$ on the variables
indicates respectively the variables of the first (left) and second (right) pendulum.
The simulations illustrate the behavior of the interconnected
dynamics for a sinusoidal exogenous signal  $q_1 = 1 +\gamma \sin(\pi t)$ 
and for a constant signal $q_2 = 0$,
for different initial conditions of the first pendulum (left).}
\label{fig:passivity}
\end{figure}

\subsection{Horizontal contraction}
Contraction theory shows how a differential analysis can infer global properties of the incremental dynamics from
the linearized system. It opens a number of possibilities to study the  incremental stability properties required by
questions including nonlinear regulation, tracking, observer design, and synchronization.

It is however well recognized that the contraction property is  the exception rather than the rule in most applications
because a number of system properties preclude contraction along {\it some} directions of the tangent space.
A system with conserved quantities or symmetries cannot be a contraction because no contraction is allowed
along the symmetry directions. A system with a limit cycle cannot be a contraction because no contraction is
allowed along the closed orbit of an autonomous system. \emph{Horizontal} contraction generalizes the differential analysis
to such situations by decomposing each tangent space into a vertical component where contraction is not required
and a horizontal space where contraction is required. 
The recent paper \cite{Forni2014} explores particular situations where
this local decomposition can lead to a global analysis of the behavior. 

Within the differential Lyapunov theory, weak forms of contraction can be 
easily introduced by weakening \eqref{eq:finslyap} to the inequalities
\begin{equation}
\label{eq:finslyap_horizontal}
 c_1 |\pi(x) \delta x|_x^p \leq V(x,\delta x) \leq c_2 |\pi(x) \delta x|_x^p 
\end{equation}
where, for every $x\in\calX$, $\pi(x)$ is a linear projection that maps the tangent vectors of
$T_x\calX$ into the horizontal subspace $\calH_x \subset T_x\calX$. 
In local coordinates
$\pi(x)$ is a matrix whose elements are smooth functions of $x$. 
Its columns provide a horizontal distribution spanning $\calH_x$. 
The vertical space $\calV_x$ is thus defined by the vectors $\delta x \in T_x\calX$ such that 
$\pi(x)\delta x = 0$. 

The combination of \eqref{eq:finslyap_horizontal} and \eqref{eq:finslyap_decay} 
establishes a contraction property confined to the directions spanned by the horizontal distribution.
It opens the way to the use of the differential Lyapunov theory in the presence of symmetry directions
along which no contraction is expected, \cite[Section VII]{Forni2014}. Figure \ref{fig:limit_cycle} provides
an illustration of the approach. The transversality of the horizontal space (in blue)
with respect to the motion along the limit cycle $\Omega$ (in black) allows to disregard the 
lack of contraction in the direction of the vector field of the system (in red). 
Indeed, in a small neighborhood of the limit cycle, 
the integral manifold of the horizontal distribution at $x\in \Omega$ is a 
Poincar{\'e} section (see Figure \ref{fig:poincare}).
\begin{figure}[htbp]
\centering
\includegraphics[width=0.48\columnwidth]{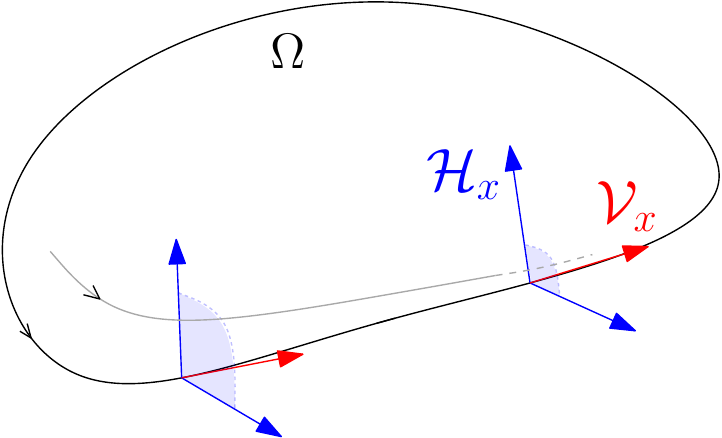}
\caption{Horizontal contraction decomposes the tangent space into a horizontal space (blue)
and a vertical space (red). Contraction is established along the horizontal directions only.}
\label{fig:limit_cycle}
\end{figure}

Several examples of weak contraction for
limit cycle analysis and synchronization can be found in
\cite{Pham2007,Russo2011a,Russo2013,Forni2014,Manchester2014}.
The property introduced in the next section is also a form of horizontal 
contraction that is relevant for the pendulum analysis.

\section{Differential positivity: local order and simple attractors}
\label{sec:positivity}

\subsection{A differential view on monotone systems}

Monotone dynamical systems  \cite{Smith1995,Hirsch1995}
are dynamical systems whose trajectories preserve some 
partial order relation $\preceq_\calK$ on the state space.
Partial orders are usually defined from cones.
Let  $\calV$ be the 
vector state-space of the system and consider a (pointed convex solid) cone 
$\calK\subseteq\calV$. For any given $x_1,x_2\in \calV$, 
the partial order $\preceq_\calK\subseteq\calV\times\calV$ satisfies
\begin{equation}
\label{eq:order_banach_X}
x_1\preceq_\calK x_2 \quad \mbox{iff} \quad x_2-x_1 \in \calK \ .
\end{equation}
From \eqref{eq:order_banach_X}, monotonicity reads as follow.
For any initial time $t_0$, 
any pair of trajectories $x_1(\cdot), x_2(\cdot)$ of a monotone system
satisfies
\begin{equation}
\label{eq:monotonicity}
 x_1(t_0) \preceq x_2(t_0) 
 \ \Rightarrow\
 x_1(t) \preceq x_2(t) \quad \forall t\geq t_0
\end{equation}
Through the introduction of a partial order relation on the input space, 
the notion of monotonicity easily extends to open systems,
as illustrated in \cite{Angeli2003}.

Monotone systems
include the class of cooperative and competitive systems
\cite{Hirsch2003,Piccardi2002} and play a fundamental role 
in chemical and biological applications
 \cite{Angeli2004,DeLeenheer2004,DeLeenheer2007,Sontag2007,Angeli2008,Angeli2012}.
They enjoy important convergence properties \cite{Smith1995,Hirsch1988,Mierczynski1995,Angeli2004a,Angeli2006a,Banaji2010} 
and interesting interconnections properties  \cite{Angeli2003,Angeli2004b,Enciso2005}.

A crucial observation is that monotonicity of a system
$\dot{x} = f(x)$ is equivalent to the \emph{positivity} of the
linearized dynamics $\dot{\delta x} = \partial f(x)\delta x$. 
Positivity is intended here in the sense of cone invariance \cite{Bushell1973}:
for any initial time $t_0$ the trajectories $\delta x(\cdot)$ of the 
linearized positive dynamics satisfy the implication
\begin{equation}
\label{eq:positivity2}
\delta x(t_0) \in \calK \  \Rightarrow  \ \delta x(t) \in \calK \quad \forall t\geq t_0 \ .
\end{equation}

An intuitive explanation of the connection between positivity
and monotonicity follows from the analysis of the mismatch
between infinitesimally neighboring solutions $ x_2(\cdot) := x_1(\cdot) + \delta x(\cdot)$
of the nonlinear dynamics $\dot{x} = f(x)$,
where
$\delta x(\cdot)$ is driven by the linearized dynamics
$\dot{\delta x}(t) = \partial_x f(x_1(t))\delta x(t)$.
The combination of 
\eqref{eq:order_banach_X} and \eqref{eq:monotonicity} 
gives 
\begin{equation}
\label{eq:monotonicity2}
 x_2(t_0) - x_1(t_0) \in \calK
 \ \Rightarrow\ 
 x_2(t) - x_1(t) \in \calK \quad \forall t\geq t_0 \ 
\end{equation}
which leads to \eqref{eq:positivity2}
because of the identity $\delta x(\cdot) =  x_2(\cdot) - x_1(\cdot)$.
The route from \eqref{eq:monotonicity} to \eqref{eq:positivity2} is at the core 
of the equivalence between closed cooperative systems 
and the Kamke condition \cite[Chapter 3]{Smith1995},
and of the equivalence between open cooperative systems and the notion of
incrementally positive systems introduced in \cite[Section VIII]{Angeli2003}.

We anticipate that a suitably extended notion of positivity,
detailed in the next section,
is the source of several convergence properties of monotone systems.
Again, a property of the linearization (positivity) 
underlies a property of the nonlinear system (monotonicity)

\subsection{Positive linearizations}

Positive systems are linear behaviors that leave a cone invariant \cite{Bushell1973}.
Rephrasing \eqref{eq:positivity2}, 
the linear system $\dot{x} = Ax$ is positive
with respect to a cone $\calK$ if
\begin{equation}
\label{eq:positivity}
 e^{At}\calK \subseteq \calK \qquad \forall t > 0 \ ,
\end{equation}
where  $e^{At}\calK := \{ e^{At}x\,|\, x\in \calK\}$.
Positive systems have a rich history because positivity strongly restricts the behavior
of the system, as established by the Perron-Frobenius theory: under mild extra assumptions ensuring that the
the transition matrix $e^{At}$ maps the boundary of the cone into the interior,
any trajectory $e^{At}x$, $x\in \calK$, converges asymptotically to a one dimensional 
subspace spanned by the eigenvector associated to
the (real) eigenvalue of largest real part of the state matrix $A$, \cite{Bushell1973}.
This convergence follows from the fact that
positive systems enjoy a  \emph{projective contraction} property
which has been exploited in a number of applications,
ranging from stabilization \cite{Willems1976, Muratori1991,Farina2000,DeLeenheer2001,Knorn2009, Roszak2009}
to observer design \cite{Hardin2007,Bonnabel2011}, and to distributed control \cite{Moreau2004, Rantzer2012, Sepulchre2010}.

For nonlinear dynamics, the crucial observation is that positivity of the linearization
strongly restricts also the nonlinear behaviors. For dynamics on manifolds $\calX$, 
positivity must be intended in a generalized sense, compatible with the fact that
the prolonged system $\dot{x} = f(x)$, $\dot{\delta x} = \partial f(x) \delta x$
lives in the tangent bundle $T\calX$. The cone of linear positivity becomes a
(smooth) cone field given by a (pointed convex solid) cone $\calK(x) \subseteq T_x\calX$  attached to each $x$.
The nonlinear dynamics is \emph{differentially positive} if the cone field is invariant
along the trajectories of the (prolonged) system,
that is,
\begin{equation}
\label{eq:differential_positivity}
\partial \psi_t(x) \calK(x) \subseteq \calK(\psi_t(x)) \qquad \forall x\in \calX,\forall t\geq 0
\end{equation}
where $\psi_t(x)$ denotes the flow of $\dot{x} = f(x)$ at time $t$ from the initial condition $x$,
and $\partial \psi_t(x)$ denotes the differential  $\partial_x \psi_t(x)$ computed at $x$,
\cite[Section 5]{Forni2014a_ver1}.
Note that for any initial condition $(x,\delta x) \in T\calX$ the pair
$(\psi_t(x), \partial \psi_t(x)\delta x)$ is a trajectory of the prolonged system.
Differential positivity \eqref{eq:differential_positivity} reduces to positivity \eqref{eq:positivity}
on linear dynamics and constant cone fields.

Figure \ref{fig:phase_portraits} illustrates three different phase portraits of differentially positive systems.
One of the phase portraits is represented in two different set of coordinates.
The systems in Figure \ref{fig:phase_portraits}.I and Figure \ref{fig:phase_portraits}.II
are differentially positive with respect to a \emph{constant}
cone field on a \emph{vector space}. The system on the left is a linear positive system. The
one on the right is a monotone system whose partial order relation $\preceq$ is the usual element-wise order.
Indeed, every differentially positive system with respect to a \emph{constant} cone field on a \emph{vector space}
is a monotone system with respect to the order $x \preceq_{\calK} y \mbox{ iff } y-x \in \calK$.
The harmonic oscillator in
Figure \ref{fig:phase_portraits}.III is neither a positive
system, nor a monotone system, but it is a differentially positive system with respect to a \emph{non constant}
cone field rotating with the flow. In polar coordinates, that is, on the \emph{nonlinear} space
$\real_+ \times \mathbb{S}$, the coordinate representation of the cone field in each tangent space is constant.

\begin{figure}[htbp]
\centering
\includegraphics[width=0.98\columnwidth]{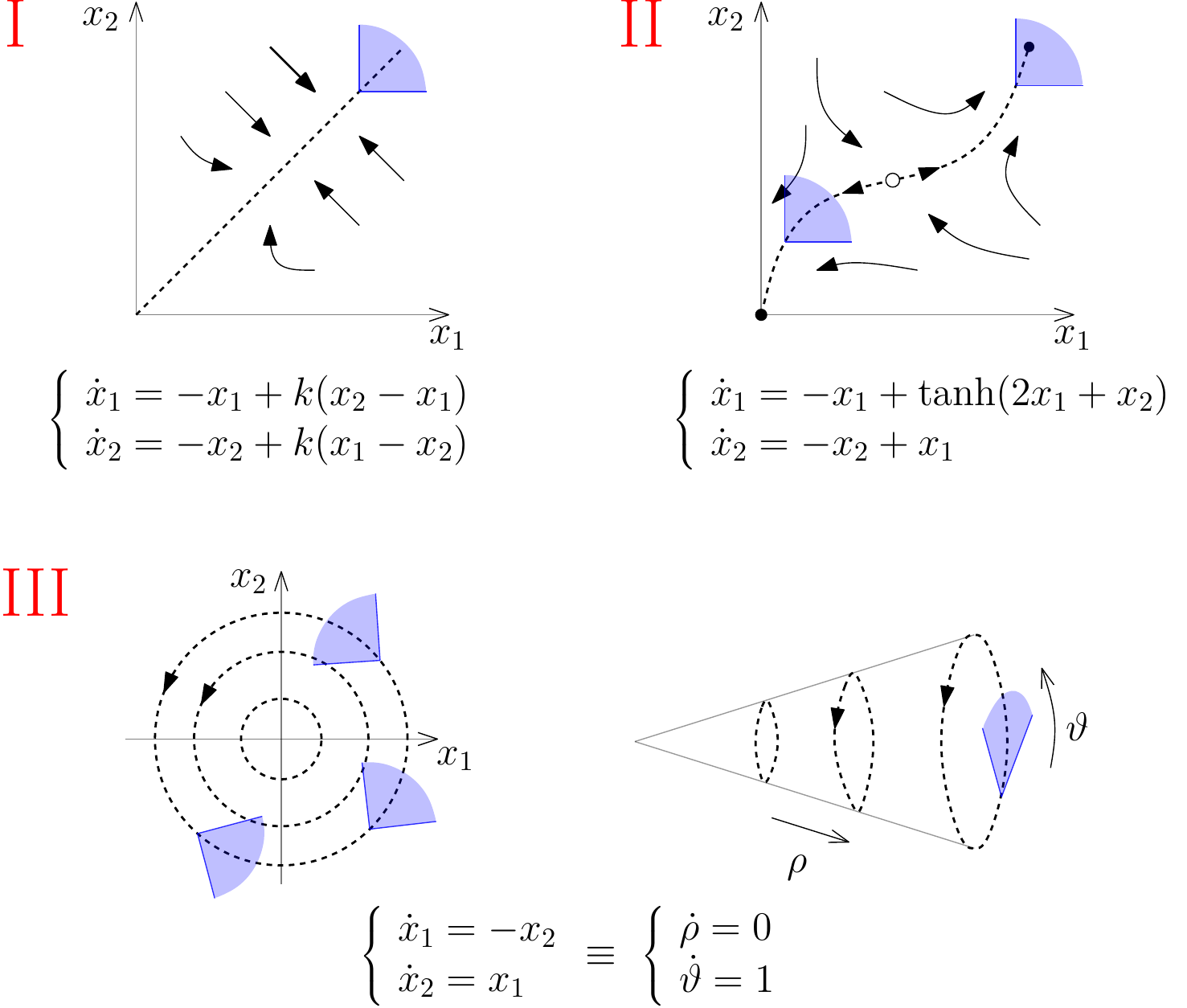}
\caption{The phase portraits of three different planar differentially positive systems.
(I) a linear consensus model.
(II) a monotone bistable model.
(III) the harmonic oscillator.
}
\label{fig:phase_portraits}
\end{figure}

Differentially positive systems inherit many properties of positivity,
under some mild extra conditions ensuring that the
the differential $\partial\psi_T(x)$ maps uniformly the boundary of the cone at $x$
into the interior of the cone $\calK(\psi_T(x))$ for some
$T> 0$ (see the notion of uniform strict differential positivity in \cite{Forni2014a_ver1}).
In particular, the projective contraction of positive systems extends
to differentially positive systems, leading
to the definition of the so called  Perron-Frobenius vector field $w(x)$ \cite[Theorem 2]{Forni2014a_ver1},
a continuous vector field direct generalization of the Perron-Frobenius dominant eigenvector of linear positivity.
Indeed, consider the distribution $\calW(x) := \{\lambda w(x)\,|\, \lambda \in \real\} \subset T_x\calX$ 
spanned by the Perron-Frobenius vector field $w(x)$. For 
any trajectory $\psi_t(x)$, the  distribution spanned by
$\calW(\psi_t(x))$ is an attractor for the linearized dynamics along the trajectory
\cite[Theorem 1]{Forni2014a_ver1}, that is,
for any $\delta x\in \calK(x)$, 
\begin{equation}
\label{eq:diff+_contraction}
\partial \psi_t(x) \delta x \to \calW(\psi_t(x)) \mbox{ as } t\to\infty \ .
\end{equation}

The identity $f(\psi_t(x)) = \partial \psi_t(x) f(x)$, 
\eqref{eq:diff+_contraction} guarantees that 
if $f(x)\in \calK(x)$ then the system vector field along $\psi_t(x)$,
satisfies 
\begin{equation}
\label{eq:diff+_contraction2}
f(\psi_t(x)) \to \calW(\psi_t(x)) \mbox{ as } t\to\infty \ .
\end{equation}

The reader will immediately recognize that the asymptotic alignment of the vector field to the 
Perron-Frobenius vector field must constrain the steady-state behavior
of differentially positive systems. In fact, exploiting \eqref{eq:diff+_contraction} and \eqref{eq:diff+_contraction2}
\cite[Theorem 3]{Forni2014a_ver1} establishes that
the limit behavior of a differentially positive system is either
described by integral curves of the Perron-Frobenius vector field, or
it is a pathological behavior, where the motion is transversal to the Perron-Frobenius
vector field, leading possibly to chaotic attractors. Precisely,
for every $x\in \calX$,
the $\omega$-limit set $\omega(x)$ satisfies
one of the following two properties:
\begin{itemize}
\item[(i)] The vector field $f(z)$ is aligned with the Perron-Frobenius vector field $w(z)$
for each $z\in \omega(x)$, and $\omega(x)$ is either a fixed point
or a limit cycle or a set of fixed points and connecting arcs;
\item[(ii)] The vector field $f(z)$ is nowhere aligned with the
Perron-Frobenius vector field $w(z)$ for each $z \in \omega(x)$,
and either
$
\liminf\nolimits\limits_{t\to\infty}
|\partial \psi_t(z)w(z)|_{\psi_t(z)} = \infty
$
or
$\lim\nolimits\limits_{t\to\infty} f(\psi_t(z)) = 0$.
\end{itemize}

The dichotomy of the limit behaviors has interesting implications
(see \cite[Section VII]{Forni2014a_ver1}). For example, the characterization above allows to show that
the trajectories of a differentially positive system
with constant cone field on a vector space converge from almost every initial condition to a fixed point,
indeed recovering the well-known property of almost global convergence of (strict) monotone dynamics,
\cite{Hirsch1988,Smith1995}.
Another interesting implication concerns limit cycles analysis.
Any compact forward invariant region $\calC \subseteq \calX$
that does not contain fixed points, and such that
$f(x)$ belongs to the interior of $\calK_\calX(x)$ for any $x\in \calC$,
necessarily contains a unique attractive periodic orbit, \cite[Corollary 2]{Forni2014a_ver1}.
See Figure \ref{fig:PB} for an illustration.
The result shows the potential of differential positivity
for the analysis of limit cycles in possibly high dimensional spaces.

\begin{figure}[htbp]
\centering
\includegraphics[width=0.6\columnwidth]{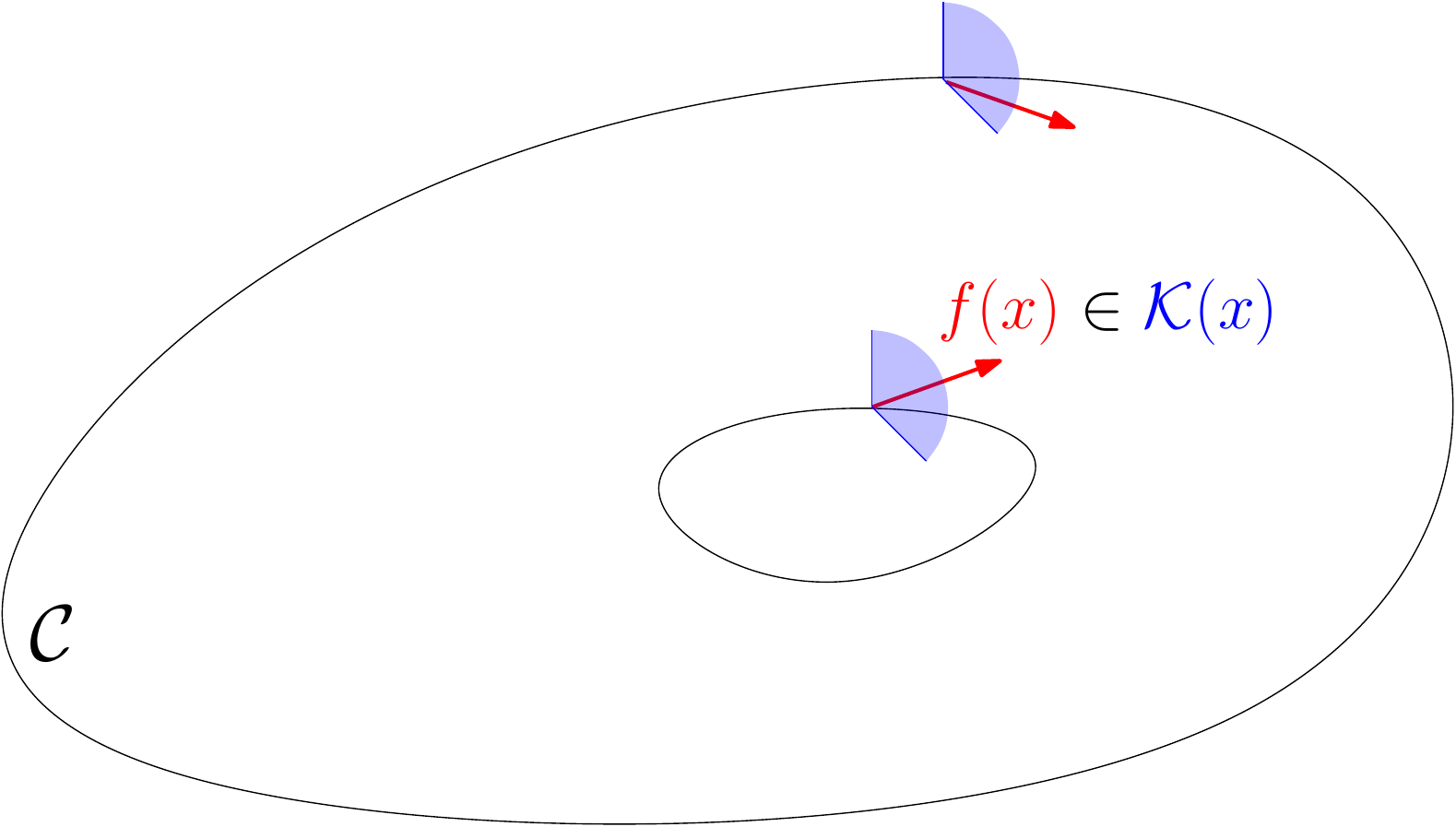}
\caption{(Strict) Differential positivity guarantees the existence of
an isolated and attractive limit cycle in every
compact forward invariant region $\calC$ that does not contain any fixed point
and satisfies the condition that $f(x)$ belongs to the interior of $\calK_\calX(x)$ for any $x\in \calC$.
}
\label{fig:PB}
\end{figure}

\subsection{Differential positivity of the nonlinear pendulum}

For values of the damping $k$ greater than $2$, the
(strict) differential positivity of the pendulum can be established
by looking at the state matrix $A(\vartheta,k)$ in \eqref{eq:closed_linearization}.
The invariant cone field reads
\begin{equation}
 \calK(\vartheta,v) := \{(\delta \vartheta, \delta v)\in T_{(\vartheta,v)}\calX \,|\, \delta \vartheta \geq 0, \delta \vartheta+\delta v \geq 0   \}  \ ,
\end{equation}
represented by the shaded region in Figure \ref{fig:pendulum_A}. The invariance follows
from the observation that
for any $\delta x = [ \ \delta \vartheta \ \  \delta v \ ]^T$ on the boundary of the cone, the
vector field of the linearized dynamics $A(\vartheta,k) \delta x$ is oriented towards the
interior of the cone for any value of $\vartheta$, as represented by the black arrows
attached to the boundary of the cone in Figure \ref{fig:pendulum_A}.
The blue and the red lines in Figure \ref{fig:pendulum_A}
show the direction of the eigenvectors of
$A(\vartheta,4)$ (left) - $A(\vartheta,3)$ (center) - $A(\vartheta,2)$ (right),
for sampled values of $\vartheta \in \mathbb{S}$.
The red eigenvectors are
related to the largest eigenvalues and play the role of attractors for the linearized dynamics.
The projective contraction holds for $k>2$ and it is lost at $k=2$, for which the state matrix
$A(0,2)$ has two eigenvalues in  $-1$ that makes the positivity of the linearized system on the
equilibrium at $0$ (for $u = 0$) non strict.

\begin{figure}[htbp]
\centering
\includegraphics[width=0.32\columnwidth]{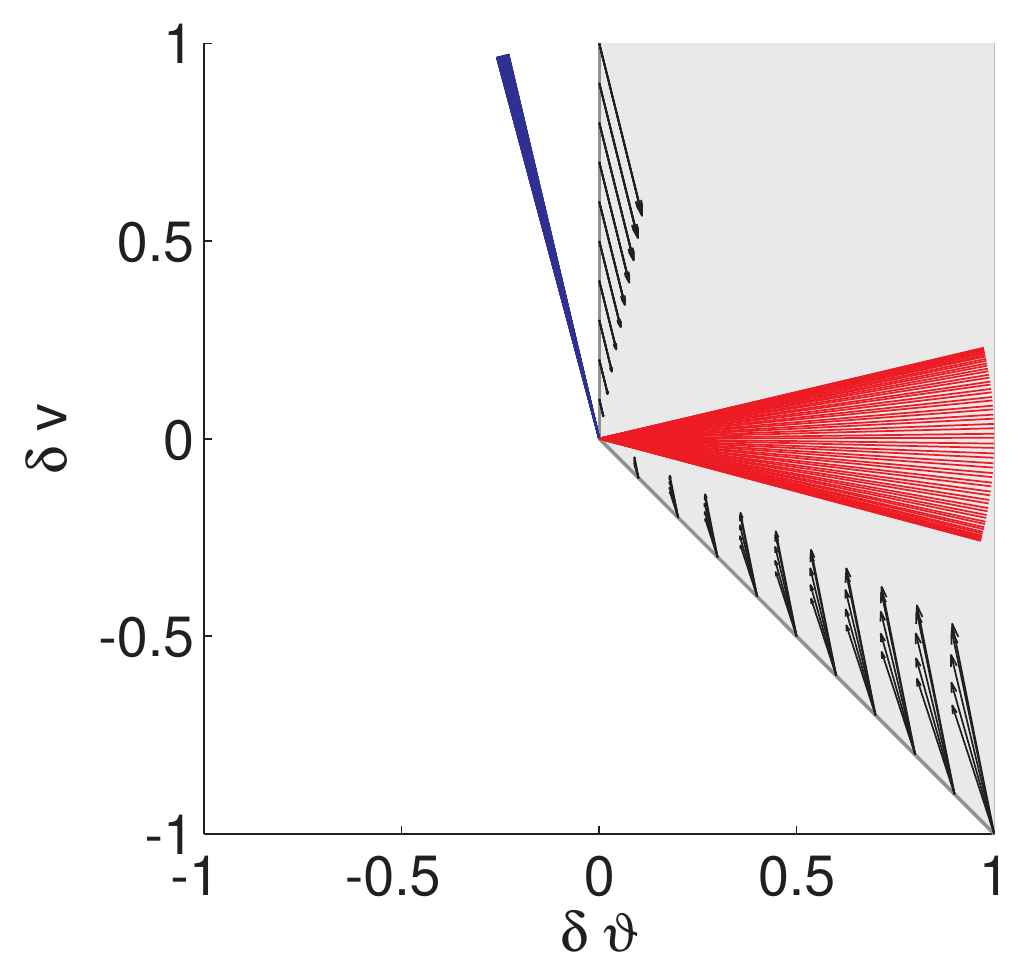}
\includegraphics[width=0.32\columnwidth]{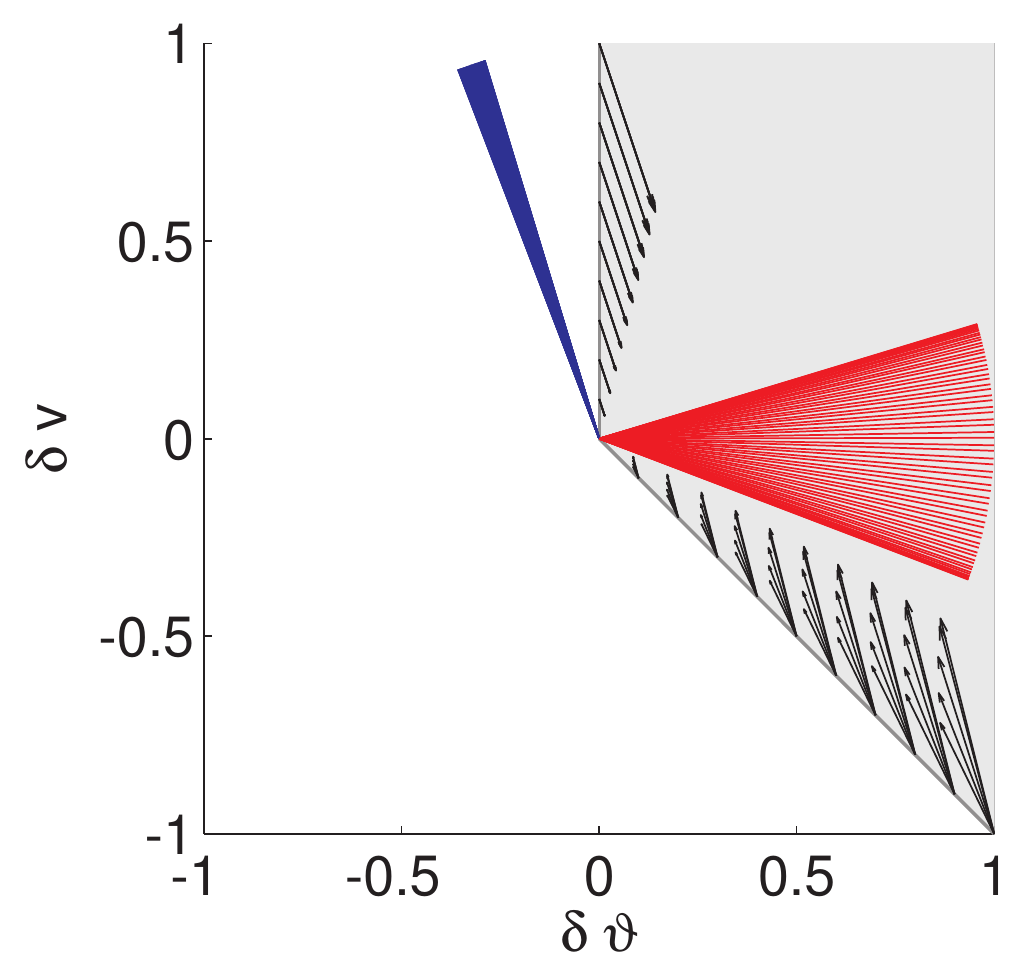}
\includegraphics[width=0.32\columnwidth]{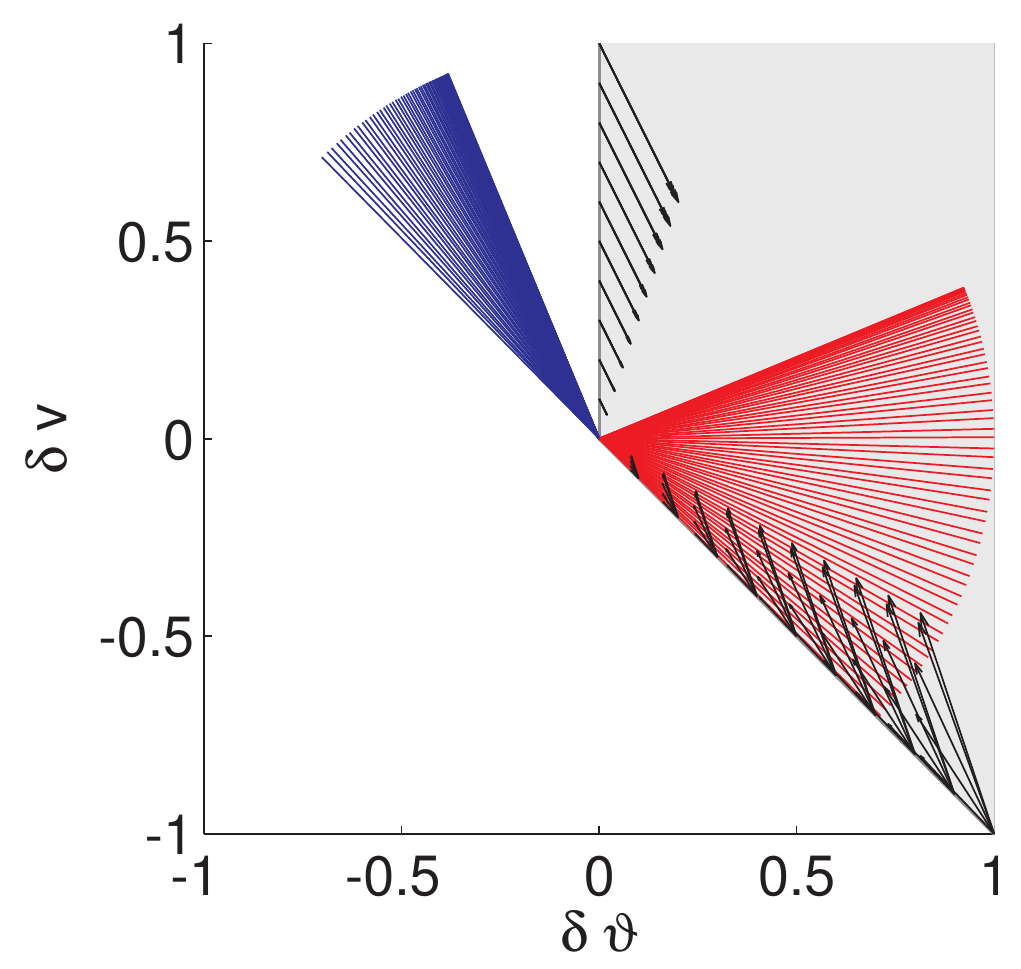}
\caption{$\calK(\vartheta,v)$ is shaded in gray.
The two eigenvector of $A(\vartheta,k)$ for a selection $\vartheta \in \mathbb{S}$
are represented in red and blue.
Red eigenvectors - largest eigenvalues. Blue eigenvectors - smallest eigenvalues.
Left figure: $k=4$. Center figure: $k=3$. Right figure: $k=2$.
}
\label{fig:pendulum_A}
\end{figure}

For $k>2$ the trajectories of the pendulum are bounded. In particular,
the kinetic energy $E := \frac{v^2}{2}$ satisfies
$\dot{E} = -kv^2 + v(u - \sin(\vartheta)) \leq (|u|+1-k|v|)|v|  < 0$, which
guarantees finite time convergence of the velocity component
towards  the set
$
\calV := \{v \in\real \,|\, -\rho \frac{|u|+1}{k} \leq v \leq \rho \frac{|u|+1}{k}\}
$
for any given $\rho > 1$.

The compactness of the set $\mathbb{S}\times \calV$ opens the way
to the use of the results of the previous section.
For $u = 1+ \varepsilon$, $\varepsilon > 0$,
we have that
$\dot{v} \geq  \varepsilon - kv$ which, after a transient,
guarantees that
$\dot{v} > 0$, thus eventually $\dot{\vartheta} > 0$.
Denoting by $f(\vartheta, v)$ the right-hand side in \eqref{eq:pendulum},
it follows that, after a finite amount of time,
every trajectory belongs to a forward invariant set $\calC \subseteq \mathbb{S}\times \calV$
such that $f(\vartheta,v)$ belongs to the interior of $\calK(\vartheta,v)$.
Thus, the region $\calC$ contains an isolated and attractive limit cycle.

\subsection{Differential positivity and homoclinic orbits}

Besides projective contraction, differential positivity introduces a local order on the system dynamics
that is not compatible with specific behaviors, like the existence of classes of homoclinic
orbits. In particular, it rules out the existence of homoclinic orbits like the one illustrated in Figure \ref{fig:homoclinic_simple}. 
In view of the discussion of Section III.C, this means that differential positivity rules out a main route to 
complex attractors by imposing locally a partial order on solutions.
The intuitive explanation is based on the fact that the linearization along a specific trajectory is
an approximation of the mismatch between the specific trajectory and the neighboring ones. Within this
interpretation, the invariance property of the cone field enforces on the system state space a local order
relation that must be preserved among neighboring trajectories. 
For instance, on vector spaces for simplicity,
consider an homoclinic orbit like the one described by the dashed line in Figure \ref{fig:homoclinic2}.
The red arrows represent the direction of the Perron-Frobenius vector field.
We show that such a homoclinic orbit is not compatible with differential positivity.
Take two initial conditions $x$ and $x+\varepsilon \delta x$,
$\varepsilon>0$ small, such that 
$x$ and $x+\varepsilon \delta x$ belong to the unstable manifold
of the saddle point, in an infinitesimal neighborhood $\calU$ of the saddle point.
For $\calU$ sufficiently small, by continuity, $\delta x \in \calK(x)$ 
since the Perron-Frobenius vector field at the saddle point is tangent to the unstable manifold of the saddle.  
For $\varepsilon$ sufficiently small, 
the trajectories $\psi_t(x)$ and $\psi_t(x+\varepsilon \delta x)$
satisfy $\psi_t(x+\varepsilon \delta x) - \psi_t(x) \simeq \varepsilon \partial \psi_t(x) \delta x \in \calK(\psi_t(x))$ for $t\geq 0$.
Moreover, because of the homoclinic orbit, for some $t>0$, 
$\psi_t(x)$ and $\psi_t(x+\varepsilon \delta x)$ 
return to the saddle point along the stable manifold, thus necessarily breaking the relation
$\psi_t(x+\varepsilon \delta x) - \psi_t(x)  \in \calK(\psi_t(x))$.
The invariance property on the cone field necessarily fails. 

\cite[Corollary 3]{Forni2014a_ver1} claims that 
under (strict) differential positivity, any homoclinic orbit of a hyperbolic fixed point
cannot be tangent to the Perron-Frobenius vector field $w(x)$
for any $x$ on the orbit.
The claim is well illustrated in Figure \ref{fig:homoclinic2}.
The stable and unstable manifolds of the saddle have
dimension 1 and 2, respectively. The
homoclinic orbit on the right part of the figure (dashed line)
is ruled out by the local order at the saddle point.  Rephrasing the argument above, along
the whole dashed orbit the vector field $f(x)$ must be parallel along the whole Perron-Frobenius
vector field $w(x)$, which violates continuity of the Perron-Frobenius vector field
at the saddle point.
The limit set given by the homoclinic orbit on the
left part of the figure (solid line) is instead compatible
with differential positivity, but the Perron-Frobenius vector
field is necessarily nowhere tangent to the curve.

\begin{figure}[htbp]
\centering
\includegraphics[width=0.5\columnwidth]{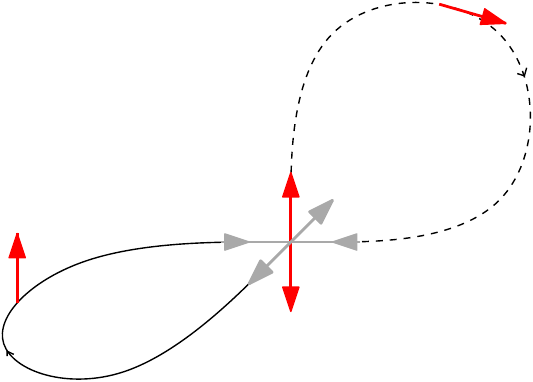}
\caption{
Two examples of homoclinic orbits.
The dashed one is ruled out by (strict) differential positivity.
The left one is compatible with differential positivity. The red
vectors represent the direction of the Perron-Frobenius vector field.}
\label{fig:homoclinic2}
\end{figure}

This analysis has a direct consequence on the pendulum example.
Looking at Figure \ref{fig:critical_damping},
the differential positivity of the pendulum  for $k\geq 2$ cannot be
extended to values of the damping $k<k_c$ because of the presence of a
homoclinic bifurcation for suitably selected values of the torque.
Still, differential positivity might hold within the invariant subregions of the
system state space separated by the homoclinic orbit.

\section{Conclusion}
Differential analysis aims at exploiting the (local) properties of linearized
dynamics to infer (global) properties of nonlinear behaviors. It is especially
relevant for the analysis of nonlinear models defined by nonlinear vector fields
on nonlinear spaces. The tutorial paper has illustrated on the nonlinear pendulum
example reasons why several nonlinear control problems require an analysis of the incremental dynamics
and the potential of differential analysis to address such questions. Emphasis was
put on recent developments by the authors in differential analysis \cite{Forni2014, Forni2014a_ver1}.
Horizontal contraction and differential positivity illustrate the potential of a differential
analysis beyond the global analysis of an equilibrium solution. It is hoped that
the insight provided by a differential analysis in an archetype model such as the
 nonlinear pendulum  will stimulate the potential relevance of this approach  in more challenging control
applications.

\end{document}